\documentclass[11pt]{article}
\usepackage[pdftex]{graphicx,color} 
\usepackage{jheppub}
\usepackage{amsmath}
\usepackage{amssymb}
\usepackage{comment}
\usepackage{multirow}
\usepackage{mathtools}
\usepackage{dsdshorthand}
\newcommand{\bea}{\begin{equation}\begin{aligned}}
\newcommand{\eea}[1]{\label{#1}\end{aligned}\end{equation}}
\newcommand{\beq}{\begin{equation}}
\newcommand{\eeq}{\end{equation}}
\newcommand   \xb  {\bar{x}}
\newcommand   \zb  {\bar{z}}

\def\O{{\mathcal O}}
\DeclareMathOperator\Disc{Disc}

\newcommand{\es}[2] {\begin{equation} \label{#1} \begin{split} #2 \end{split} \end{equation}}

\usepackage{tikz}
\usetikzlibrary{arrows,calc,shapes,decorations.pathmorphing,decorations.markings}
\tikzset{
>=stealth',
help lines/.style={dashed, thick},
axis/.style={<->},
important line/.style={thick},
connection/.style={thick, dotted},
  cross/.style={
    cross out,
    draw=black, 
    minimum size=5pt, 
    inner sep=0pt,
    outer sep=0pt
  },
->-/.style={decoration={
  markings,
  mark=at position #1 with {\arrow{>}}},postaction={decorate}}
}

\title{AdS Virasoro-Shapiro from dispersive sum rules}

\author{Luis F. Alday, Tobias Hansen and Joao A. Silva}

\affiliation{Mathematical Institute, University of Oxford,
Woodstock Road, Oxford, OX2 6GG, UK}

\abstract{
We consider the four-point correlator of the stress-energy tensor in ${\cal N}=4$ SYM, to leading order in inverse powers of the central charge, but including all order corrections in $1/\lambda$. This corresponds to the AdS version of the Virasoro-Shapiro amplitude to all orders in the small $\alpha'$/low energy expansion. Using dispersion relations in Mellin space, we derive an infinite set of sum rules. These sum rules strongly constrain the form of the amplitude, and determine all coefficients in the low energy expansion in terms of the CFT data for heavy string operators, in principle available from integrability. For the first set of corrections to the flat space amplitude we find a unique solution consistent with the results from integrability and localisation.}

\emailAdd{alday@maths.ox.ac.uk, tobias.hansen@maths.ox.ac.uk, joao.silva@maths.ox.ac.uk}

\begin{document}
\maketitle

\section{Introduction}

The AdS/CFT duality gives us a tool to compute string scattering amplitudes on $AdS_5 \times S^5$ by computing correlators in the dual ${\cal N}=4$ SYM theory at the boundary. According to the AdS/CFT dictionary the genus expansion in powers of $g_s$, on the amplitude side, corresponds to an expansion in inverse powers of the central charge $c$, while the low energy expansion, in powers of $\alpha'$, corresponds to an expansion in inverse powers of the 't Hooft coupling $\lambda$. 

In this paper we consider the simplest four-point amplitude, corresponding to the scattering of four graviton states at tree level. In flat space this corresponds to the Virasoro-Shapiro amplitude. In $AdS_5 \times S^5$ this corresponds to the correlator of four stress-tensor multiplets, to leading non-trivial order in a $1/c$ expansion. The best language to express this correlator is Mellin space \cite{Penedones:2010ue}. To leading order in $1/c$ and in a low energy expansion it takes the form
\begin{equation}
M(s,t) = \frac{8}{(s-2)(t-2)(u-2)} + \sum\limits_{a,b=0}^\infty  \frac{\Gamma(2a+3b+6)}{8^{a+b}\lambda^{\frac32 + a + \frac32 b} } \sigma_2^a \sigma_3^b
\left(  \a_{a,b} + \frac{\b_{a,b} }{\lambda^{\frac{1}{2}}}  + \frac{ \gamma_{a,b}}{\lambda}
 + \cdots \right)
\end{equation}
where $s+t+u=4$, $\sigma_2=s^2+t^2+u^2$ and $\sigma_3=s t u$, and we have suppressed an overall $1/c$. This is the AdS analogue of the Virasoro-Shapiro amplitude, and will be the main object of study of this paper. The first term in the low energy expansion corresponds to the correlator in the supergravity approximation. In addition we have an infinite tower of stringy corrections, given by symmetric polynomials of higher and higher degree. Over the last years a lot of effort has been devoted to fix these coefficients. The leading order coefficients $\alpha_{a,b}$ are fixed by comparison to the flat space Virasoro Shapiro amplitude, for instance $\alpha_{a,0}=\zeta(3+2a)$. In addition, supersymmetric localisation imposes two extra constraints at each order in $1/\lambda$, see  \cite{Binder:2019jwn,Chester:2020dja}. 

There are reasons to believe that the full expression for $M(s,t)$ ought to be very complicated. In particular it should encode complete information about the spectrum and OPE-coefficients of heavy string states, dual to single trace short operators. On the other hand this information should be available, at least in principle, from integrability results. A natural question is how these results constrain the terms in the above expansion. 

In this paper we will use a fact that so far hasn't been used in the context of ${\cal N}=4$ SYM. Namely, the Mellin amplitude under consideration satisfies
\cite{Maldacena:2015waa, Caron-Huot:2021rmr}
\begin{align}\label{BoundChaos_intro}
\lim_{s \rightarrow i \infty} |M(s, t)| \leq \frac{1}{|s|^2}, ~~ \text{Re} (t) < 2 \,.
\end{align}
Note that this bound would be violated by any finite number of contact terms. Following \cite{Penedones:2019tng} this simple fact leads to a set of dispersive sum rules (see also \cite{Carmi:2019cub, Caron-Huot:2020adz, Caron-Huot:2021enk}) relating the coefficients in the low energy expansion to the CFT-data for heavy string operators. These equations have a relatively rigid structure and constrain not only the low energy coefficients but also the CFT-data that enters the equations. Under some mild assumptions we are able to compute the first layer of corrections to the flat space Virasoro-Shapiro amplitude
\bea
\beta_{a,0}={}&2 \zeta (2 a+1)\zeta (3) -2 \zeta (2 a+1,3)-(2 a+1) \zeta (2 a+2,2)\\
&+(a-1) (2 a+7) \zeta (2 a+3,1)-\frac{1}{3} a \left(4 a^2+12 a-1\right) \zeta (2 a+4)\,.
\eea{beta_a0_result_intro}
which agrees with the localisation constraints for $a=0,1$. Higher order coefficients $\beta_{a,b}$  are also highly constrained in their form and can be completely determined once the CFT-data becomes available from integrability. 

This paper is organised as follows. In section 2 we describe some background material regarding  the stress tensor four-point function in $\mathcal{N}=4$ SYM. In section 3 we derive the Mellin dispersive sum rules, while in section 4 we analyse the consequences of the flat space limit for the CFT-data of heavy string operators. In section 5 we go into AdS, focusing on the first layer of  corrections to the flat space Virasoro-Shapiro amplitude. Finally in section 6 we end up with some conclusions, while many technical details are included in the appendices. 

\section{$\mathcal{N}=4$ stress tensor four-point function}

\subsection{Basics}

The main object of study in this paper will be the correlator of the component $\cO_{2 \, IJ}(\vec x)$ of the stress-tensor multiplet, which is a dimension $2$ scalar in the ${\bf 20}'$ of the R-symmetry group $SU(4)_R \cong SO(6)_R$. We will contract its two $SO(6)_R$ indices with polarisation vectors $Y^I$, $I=1,\ldots,6$ satisfying $Y\cdot Y = 0$ and write the correlator as
\es{2222}{
 & \langle \cO_2(\vec x_1,Y_1) \cO_2(\vec x_2,Y_2) \cO_2(\vec x_3,Y_3) \cO_2(\vec x_4,Y_4) \rangle = \frac{Y^2_{12}Y^2_{34}}{{x}_{12}^4 {x}_{34}^{4}} \mathcal{S}(U,V;\sigma,\tau)\,,
    }
where we define the cross-ratios
 \es{uvsigmatauDefs}{
  U \equiv \frac{{x}_{12}^2 {x}_{34}^2}{{x}_{13}^2 {x}_{24}^2} \,, \qquad
   V \equiv \frac{{x}_{14}^2 {x}_{23}^2}{{x}_{13}^2 {x}_{24}^2}  \,, \qquad
   \sigma\equiv\frac{(Y_1\cdot Y_3)(Y_2\cdot Y_4)}{(Y_1\cdot Y_2)(Y_3\cdot Y_4)}\,,\qquad \tau\equiv\frac{(Y_1\cdot Y_4)(Y_2\cdot Y_3)}{(Y_1\cdot Y_2)(Y_3\cdot Y_4)} \,.
 }
 The constraints of superconformal symmetry imply that the correlator can be written as \cite{Dolan:2001tt}
  \es{T}{
 \mathcal{S}(U,V;\sigma,\tau)&=\mathcal{S}_\text{free}(U,V;\sigma,\tau)+\Theta(U,V;\sigma,\tau)\mathcal{T}(U,V)\,,\\
 \Theta(U,V;\sigma,\tau)&\equiv\tau+[1-\sigma-\tau]V+\tau[\tau-1-\sigma]U+\sigma[\sigma-1-\tau]UV+\sigma V^2+\sigma\tau U^2\,,
 }
where $\mathcal{S}_\text{free}(U,V;\sigma,\tau)$ is the free theory correlator
\es{free}{
\mathcal{S}_\text{free}(U,V;\sigma,\tau)=1+U^2\sigma^2+\frac{U^2}{V^2}\tau^2+\frac{1}{c}\left(U\sigma+\frac UV\tau+\frac{U^2}{V}\sigma\tau\right)\,,
}
and we are interested in the reduced correlator $\mathcal{T}(U,V)$ which satisfies the crossing relation
\beq
\cT\left(U,V \right) =  \cT\left(1/U,V/U \right) =  \frac{1}{V^2} \cT\left(U/V,1/V \right)\,,
\label{crossing_symmetry_T}
\eeq
and can be expanded in blocks as
 \es{Texp}{
 \cT(U,V)=U^{-2}\sum_{\tau,\ell}C^2_{\tau,\ell}G_{\tau+4,\ell}(U,V)\,,
 }
 where $G_{\tau,\ell}(U,V)$, with twist $\tau=\De-\ell$ and spin $\ell$, are 4d conformal blocks
 \es{4dblock}{
 G_{\tau,\ell}(U,V) &=\frac{z\bar z}{z-\bar z}(k_{\tau+2\ell}(z)k_{\tau-2}(\bar z)-k_{\tau+2\ell}(\bar z)k_{\tau-2}( z))\,,\\
k_h(z)&\equiv z^{\frac h2}{}_2F_1(h/2,h/2,h,z)\,.
 }

\subsection{Mellin amplitude}

For our purposes we consider the Mellin transform $M(s, t)$ of the reduced correlator
 \es{MellinDef}{
  \cT(U, V)
   = \int_{-i \infty}^{i \infty} \frac{ds\, dt}{(4 \pi i)^2} U^{\frac s2} V^{\frac t2 - 2}
    \Gamma \bigg[2 - \frac s2 \bigg]^2 \Gamma \bigg[2 - \frac t2 \bigg]^2 \Gamma \bigg[2 - \frac u2 \bigg]^2
    M(s, t) \,,
 } 
 where $s+t+u=4$ and crossing symmetry \eqref{crossing_symmetry_T} implies $M(s,t) = M(t,s) = M(s,u)$.

Exchanged particles in the s-channel ($12 \to 34$) lead to simple poles for $M(s, t)$ at $s = \tau + 2m$, $m \in \mathbb{N}_0$, with residues given by
\begin{align}
M(s, t) \approx \frac{C^2_{\tau, \ell} \mathcal{Q}_{\ell, m}^{\tau +4, d=4}(t-4) }{s-\tau - 2m},
\end{align}
where $\mathcal{Q}_{\ell, m}^{\tau , d}(t)$ is a function related to Mack polynomials that we define in appendix \ref{app:Mack appendix}.

In this paper we will consider the Mellin amplitude at large $c$, to leading order in $1/c$. In that case, it is possible to bound its asymptotic behaviour at infinity through the bound on chaos \cite{Maldacena:2015waa, Caron-Huot:2021rmr}
\begin{align}\label{BoundChaos}
\lim_{s \rightarrow i \infty} |M(s, t)| \leq \frac{1}{|s|^2}, ~~ \text{Re} (t) < 2 \,.
\end{align}
This simple fact will be crucial in the next section. 

\subsection{Strong coupling expansion}

To leading order in $1/c$ the Mellin amplitude describes tree level string theory in AdS. We will furthermore consider an expansion around large $\lambda$, where the Mellin amplitude is given by the supergravity result plus a tower of stringy corrections. In Mellin space the Witten diagrams for tree level supergravity and contact diagrams (which can be used to write stringy corrections) are remarkably simple and we have the expansion
\begin{align}
M(s,t) ={}&
\frac{8}{(s-2)(t-2)(u-2)}
+\sum\limits_{a,b=0}^\infty  \frac{\sigma_2^a \sigma_3^b}{\lambda^{\frac32 + a + \frac32 b} } 
\left(  \tl\a_{a,b} + \frac1{\lambda^{\frac12}} \tl\b_{a,b}
 + \frac1{\lambda} \tl\gamma_{a,b}
 + O\left(\frac1{\lambda^{\frac32}}\right)\right)\nonumber\\
 ={}&
 \frac{8}{(s-2)(t-2)(u-2)}
 +\frac{1}{\lambda^{\frac32}} \tl\a_{0,0} + \frac1{\l^{2}} \tl\b_{0,0} + \frac1{\l^{\frac52}} \left(\tl\a_{1,0} \sigma_2 +\tl\gamma_{0,0} \right)
 \label{M_strong_coupling}\\
 &+ \frac1{\l^{3}} \left(\tl\a_{0,1} \sigma_3 +\tl\b_{1,0} \sigma_2 + \tl\delta_{0,0} \right)
  + \frac1{\l^{\frac72}} \big(\tl\alpha_{2,0} \sigma_2^2 + \tl\beta_{0,1} \sigma_3  +\tl\gamma_{1,0} \sigma_2+ \tl\epsilon_{0,0} \big)+ O \left( \frac1{\l^{4}}\right)
,\nonumber
\end{align}
where it is understood that we are only keeping the leading term in a $1/c$ expansion (proportional to $1/c$). Here $\tl\a_{a,b}$, $\tl\b_{a,b}$, $\tl\gamma_{a,b}$ are numerical coefficients, for which we will also define an alternative version for later convenience
\beq
\a_{a,b} = \frac{8^{a+b}}{\Gamma(2a+3b+6)} \tl\a_{a,b}\,, \qquad
\b_{a,b} = \frac{8^{a+b}}{\Gamma(2a+3b+6)} \tl\b_{a,b}\,, \quad \ldots
\eeq
In writing \eqref{M_strong_coupling} we used the fact that a symmetric polynomial in $s$, $t$ and $u$ satisfying $s+t+u=4$ can be uniquely written in the variables
\beq
\sigma_2 = s^2+t^2+u^2\,, \qquad \sigma_3 = s t u\,.
\eeq
The coefficients $\a_{a,b}$ are determined by the flat-space limit \cite{Penedones:2010ue,Fitzpatrick:2011hu,Binder:2019jwn,Chester:2020dja}
 \es{flat}{
 \frac{f(s, t)}{s t (-s-t)} = \lim_{L \to \infty} \frac{L^{6} c}{32} \int_{\kappa-i\infty}^{\kappa+ i \infty} \frac{d\alpha}{2 \pi i} \, e^\alpha \alpha^{-6} { M} \left( \frac{L^2}{2 \alpha} s, \frac{L^2}{2 \alpha} t \right) \,,
 }
 where $L$ is the AdS radius and $f(s,t)$ is the flat space Virasoro-Shapiro amplitude \cite{Virasoro:1969me,Shapiro:1970gy} divided by the supergravity amplitude
  \es{ScattAmp}{
  {\cal A}(s, t) = \cA_{R} (s, t) f(s, t) \,.
 }
More concretely we have (in terms of $\a'= L^2/\sqrt{\lambda}$)
\begin{align}
  \frac{f(s, t)}{s t (-s-t)} ={}& -\frac{(\a')^3}{64}\frac{ \Gamma \left(-\frac{1}{4} \a' s\right) \Gamma \left(-\frac{1}{4} \a' t\right) \Gamma \left(-\frac{1}{4} \a'(-s-t) \right) }{\Gamma \left(\frac{1}{4}\a' s +1\right) \Gamma \left(\frac{1}{4} \a't+1\right) \Gamma \left(\frac{1}{4} \a'(-s-t)
   +1\right) }\,,\label{flatspace_amplitude}\\
={}& \frac{1}{s t (-s-t)} \exp \left(2 \sum\limits_{n=1}^{\oo}\frac{\zeta(2n+1)}{2n+1} \left( \frac{\a'}{4}  \right)^{2n+1} \left(s^{2n+1}+t^{2n+1}+(-s-t)^{2n+1}\right) \right)\,.
\nonumber
\end{align}
Using \eqref{flat} to match \eqref{M_strong_coupling} and \eqref{flatspace_amplitude}
in a large $\lambda$ expansion gives, for example
\begin{align}
\a_{a,0} ={}& \zeta(3+2a)\,,\nonumber\\
\a_{a,1} ={}& \sum\limits_{\substack{i_1,i_2=0\\i_1+i_2=a}}^a\zeta(3+2i_1) \zeta(3+2i_2)
=-2 \zeta (2 a+5,1) + (a+\tfrac32) \zeta (2 a+6)\,,\nonumber\\
\a_{a,2} ={}& \frac{2}{3}\sum\limits_{\substack{i_1,i_2,i_3=0\\i_1+i_2+i_3=a}}^a\zeta(3+2i_1) \zeta(3+2i_2) \zeta(3+2i_3) + \frac16 (a+1)(a+2) \zeta(9+2a) \label{alphas}\\
={}& \zeta (2 a+7,2)+4 \zeta (2 a+7,1,1)-(2 a+5) \zeta (2 a+8,1)+\tfrac{1}{2} (a+1) (a+3) \zeta (2 a+9)\,,\nonumber
\end{align}
written in terms of multiple zeta values of depth $k$ and weight $s_1 + \ldots + s_k$
\beq
\zeta(s_1, \ldots, s_k) = \sum\limits_{n_1>\ldots>n_k>0} \frac{1}{n_1^{s_1} \cdots n_k^{s_k}}\,. 
\label{MZV}
\eeq
At each order in $1/\lambda$ there are also two linear constraints among all the coefficients due to supersymmetric localisation \cite{Binder:2019jwn,Chester:2020dja}. This fixes the coefficients appearing at low orders in $1/\lambda$ to
\beq
\beta_{0,0}=0\,,\quad
\gamma_{0,0}=-\frac{63}{4} \zeta(5)\,,\quad
\beta_{1,0}=-2\zeta(3)^2\,,\quad
\delta_{0,0}=-168 \zeta(3)^2\,.
\eeq

\section{Mellin dispersive sum rules for $\mathcal{N}=4$ SYM}

\subsection{Basic setup}

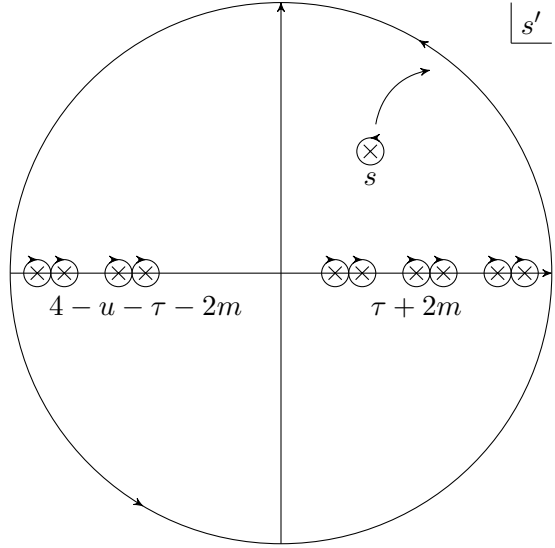
\begin{figure}
\centering
  \begin{tikzpicture}[scale=1.8]
    \coordinate (n) at (-0,2);
    \coordinate (e) at (2,0);
    \coordinate (w) at (-2,0);
    \coordinate (s) at (-0,-2);
    \coordinate (bp1) at (1,0);
    \coordinate (bp2) at (0,0);
    \draw[->] (w) --  (e) ;
    \draw[->] (s) --  (n) ;
    \node at (0.66,.9) [cross] {};  
    \draw[->-=.5] (0.66,0.8) arc (-90:270:0.1);
    \node at (0.66,0.7) [] {$s$};    
    \draw [->] (0.7,1.1) to [out=80,in=190] (1.1,1.5);    
    \node at (1.85,1.85) [] {$s'$};
    \draw[-] (2,1.7) -- (1.7,1.7);
    \draw[-] (1.7,1.7) -- (1.7,2);
    \draw[->-=.33] (2,0) arc (0:180:2);
    \draw[->-=.33] (-2,-0) arc (180:360:2);
    \node at (.4,0) [cross] {};
    \node at (.6,0) [cross] {};
    \node at (1.0,0) [cross] {};
    \node at (1.2,0) [cross] {};
    \node at (1.6,0) [cross] {};
    \node at (1.8,0) [cross] {};
    \draw[->-=.5] (0.4,-.1) arc (270:-90:0.1);
    \draw[->-=.5] (0.6,-.1) arc (270:-90:0.1); 
    \draw[->-=.5] (1.0,-.1) arc (270:-90:0.1); 
    \draw[->-=.5] (1.2,-.1) arc (270:-90:0.1); 
    \draw[->-=.5] (1.6,-.1) arc (270:-90:0.1); 
    \draw[->-=.5] (1.8,-.1) arc (270:-90:0.1); 
    \node at (-1.0,0) [cross] {};
    \node at (-1.2,0) [cross] {};
    \node at (-1.6,0) [cross] {};
    \node at (-1.8,0) [cross] {};
    \draw[->-=.5] (-1.0,-.1) arc (270:-90:0.1); 
    \draw[->-=.5] (-1.2,-.1) arc (270:-90:0.1); 
    \draw[->-=.5] (-1.6,-.1) arc (270:-90:0.1); 
    \draw[->-=.5] (-1.8,-.1) arc (270:-90:0.1); 
    \node at (1,-0.25) [] {$\tau+2m$};
    \node at (-1,-0.25) [] {$4-u-\tau-2m$}; 
  \end{tikzpicture}
\caption{Contour deformation determining the sum rules.} \label{fig:contour_deformation}
\end{figure}
We will now study the Mellin amplitude $M(s,t)$ through the use of a dispersion relation. More precisely, we will consider the quantity
\beq
\frac{M(s, t)}{ \prod_{i=1}^{q} (s- 2 -2i) (t - 2 -2i)}\,,
\eeq
where $q \in \mathbb{N}_0$ and write a dispersion relation for it. Using Cauchy's theorem,
\begin{align}
\frac{M(s, t)}{ \prod_{i=1}^{q} (s- 2 -2i) (t - 2 -2i)} = \oint_s \frac{ds'}{2 \pi i}  \frac{M(s', 4 - u - s' )}{(s'-s)\prod_{i=1}^{q} (s'- 2 -2i) (2-u-s' -2i)}\,,
\end{align}
where we keep $u=4 -s -t$ fixed and do the integral counterclockwise around a general point $s$. Afterwards we deform the contour to infinity, see figure \ref{fig:contour_deformation}. The contribution from the arc at infinity vanishes due to the bound on chaos \eqref{BoundChaos}. We pick up the physical poles at $s, t= \tau + 2 m$, $m \in \mathbb{N}_0$. There is no contribution from $s, t= 4, 6, 8, ...$ due to the Polyakov conditions (see appendix \ref{app:Polyakov conds}).

In this way we arrive at a one parameter family of sum rules $\omega_{\tau, \ell}(s, t; q)$, where $q \in \mathbb{N}_0$. The precise definition is 
\begin{align}\label{defSumRule}
\omega_{\tau, \ell}(s, t; q) &\equiv \sum_{m=0}^{\infty} \tilde{\omega}_{\tau, \ell}(s, t; q; m)\,, \\
\tilde{\omega}_{\tau, \ell}(s, t; q; m)  &= \frac{\mathcal{Q}_{\ell, m}^{\tau +4, d=4}(u-4) \left( \frac{1}{s-\tau - 2m} + \frac{1}{t-\tau - 2m} \right)}{\prod_{i=1}^{q}(2-2i-2m-u-\tau)(-2-2i+2m+\tau) }\,.
\end{align}
We then have
\begin{equation}\label{MainEquation1}
\boxed{\frac{M(s, t)}{ \prod_{i=1}^{q} (s- 2 -2i) (t - 2 -2i)} =\sum_{\O_{\tau, \ell}} C_{\O_{\tau, \ell}}^2 \omega_{\tau, \ell}(s, t; q)}\,.
\end{equation}
In the equation above we sum over all the superprimaries exchanged in the OPE of the external operators. $\tau$ and $\ell$ label the twist and spin of the superprimaries exchanged, $C_{\O_{\tau, \ell}}^2$ labels the OPE coefficient. 

The sum rule is valid to leading order in $\frac{1}{c}$ and for any t'Hooft coupling $\lambda$. We will explore its consequences at large $\lambda$. In that case, the l.h.s.\ is given by the strong coupling expansion (\ref{M_strong_coupling}). On the r.h.s.\ there are three classes of exchanged operators:\footnote{In addition there are single trace large spin operators, whose dimension go like $\Delta \sim \sqrt{\lambda} \log \ell$, but their contribution is exponentially small at large $\lambda$.}
\begin{itemize}

\item operators in short multiplets,

\item double trace operators,

\item `stringy' operators.
\end{itemize}
The stringy operators have been extensively studied in the AdS/CFT literature starting with  \cite{Gubser:1998bc}. They are single trace operators, whose conformal dimensions grow with $\lambda^{\frac{1}{4}}$. 
We will derive equations that relate the coefficients in the strong coupling expansion (\ref{M_strong_coupling}) of the Mellin amplitude with the conformal data (OPE coefficients and dimensions) of the stringy operators. This is the main result of this paper.

It is important to understand why this result is possible. The conformal block decomposition of AdS contact diagrams has been known for a long time, namely since the celebrated HPPS paper \cite{Heemskerk:2009pn}. This decomposition involves only double twist operators. It follows from \cite{Heemskerk:2009pn} that any effective field theory in AdS is compatible with the crossing equations. However, not all effective field theories in AdS are compatible with the bound on chaos (\ref{BoundChaos}). This extra information allows us to write a dispersive sum rule, which has different properties from the conformal block expansion.

\subsection{Warm-up: supergravity correlator}

We will study the sum rule (\ref{MainEquation1}) by expanding in powers of $\frac{1}{\lambda}$. As a first step, we start with the order $\frac{1}{\lambda^0}$. In that case, on the l.h.s.\ we have the supergravity correlator and on r.h.s.\ we have the short multiplets and the double trace operators with anomalous dimensions of order $\frac{1}{c}\frac{1}{\lambda^0}$:
\small
\begin{align} \label{supergravityEquation}
\frac{8}{(s-2)(t-2)(u-2)}\frac{1}{\prod_{i=1}^{q} (s- 2 -2i) (t - 2 -2i)} = \sum (\text{short multiplets}) + \sum (\text{double traces})|_{\frac{1}{c} \frac{1}{\lambda^0}}
\end{align}
\normalsize
Concerning the short multiplets, there is one superprimary for each spin $\ell=0, 2, ...$. Their OPE coefficients are equal to \cite{Beem:2013qxa, Beem:2016wfs}
\beq
C^2_\ell = -\frac{1}{c} \frac{\sqrt{\pi } 2^{-2 \ell-3} \Gamma (\ell+3)}{\Gamma \left(\ell+\frac{5}{2}\right)}\,,
\eeq
and they all have twist $2$. Since $\tilde{\omega}_{\tau=2, \ell}(s, t; q; m) = 0$ if $m =1, 2, ...$, then
\begin{align} \label{supergravityShortMultiplets}
\sum (\text{short multiplets}) = \sum_{\ell=0, 2}^{\infty} C^2_\ell \tilde{\omega}_{\tau, \ell}(s, t; q; m=0)\,.
\end{align}
The double trace operators have twists $4 + 2n$, where $n \in \mathbb{N}_0$\footnote{The double trace operators are degenerate for $n \geq 1$. This will not be relevant for us since at order $\frac{1}{c}$ their contribution to the sum rule is always proportional to their anomalous dimensions (and not to the square of the anomalous dimensions, nor to the cube, etc.)}. Their OPE coefficients $C_{n, \ell}^2$ and their anomalous dimensions $\gamma(n, \ell)$ are equal to \cite{Alday:2017xua}\footnote{This can also be deduced using dispersion relations.}
\begin{align}\label{supergravityDATA}
C_{n, \ell}^2 &= \frac{\pi  (\ell+1) 2^{-2 \ell-4 n-9} (\ell+2 n+6) \Gamma (n+3) \Gamma (\ell+n+4)}{\Gamma \left(n+\frac{5}{2}\right) \Gamma \left(\ell+n+\frac{7}{2}\right)}\,, \\
\gamma(n, \ell) &=- \frac{1}{c} \frac{(n+1) (n+2) (n+3) (n+4)}{(\ell+1) (\ell+2 n+6)}\,.
\end{align}
The functional $\omega_{\tau, \ell}(s, t; q)$ vanishes whenever $\tau = 4 + 2n$. It vanishes linearly for $n=0, 1, ..., q-1$ and it vanishes quadratically for $n=q, q+1, q+2, ...$. For this reason, only the first $q$ families of Regge trajectories of double trace operators contribute to the sum rule at order $\frac{1}{c}$. Thus,
\begin{align} \label{supergravityDoubleTraces}
\sum (\text{double traces})|_{\frac{1}{c} \frac{1}{\lambda^0}} = \sum_{\ell=0, 2}^{\infty} \sum_{n=0}^{q-1} \sum_{m=0}^{q-1} C_{n, \ell}^2  \frac{d}{d \tau} \tilde{\omega}_{\tau=4+2n, \ell}(s, t; q; m) \times \gamma(n, \ell)\,.
\end{align}
Summing the series (\ref{supergravityShortMultiplets}) and (\ref{supergravityDoubleTraces}) is a straightforward exercise that can be done for $q=0, 1, 2, ...$ and in this way one can check (\ref{supergravityEquation}) for each $q$\footnote{The strip of convergence of $\sum (\text{short multiplets})$ does not depend on $q$ and is equal to $0 < \text{Re} (u) < 2$. The strip of convergence of $\sum (\text{double traces})$ depends on $q$ and it increases with $q$. For $q=1$, it is given by $-2 < \text{Re} (u) < 2$. So, (\ref{MainEquation1}) is valid in $0 < \text{Re} (u) < 2$.}.

\subsection{Double traces}

As explained before, the double trace operators gain anomalous dimensions due to the supergravity correlator. In the same manner, each stringy correction induces anomalous dimensions proportional to it. These can be calculated using a dispersion relation. Consider a stringy correction to the Mellin amplitude $\kappa \times \sigma_2^a \sigma_3^b$. The anomalous dimensions $\tilde{\gamma}(n, \ell)$ gained by the double trace operators obey the formula
\begin{align}
\frac{ \kappa \times \sigma_2^a \sigma_3^b }{\prod_{i=1}^{q} (s- 2 -2i) (t - 2 -2i)} = \sum_{\ell=0, 2}^{2a + 3b} \sum_{n=0}^{q-1} \sum_{m=0}^{q-1} C_{n, \ell}^2  \frac{d}{d \tau} \tilde{\omega}_{\tau=4+2n, \ell}(s, t; q; m) \times \tilde{\gamma}(n, \ell)\,,
\end{align}
provided $a + b< q$. The condition $a + b< q$ ensures that we can write a dispersion relation for the contact diagram we are examining. 

For concreteness, let us work out an explicit example. Consider the contact diagram $\kappa \times \sigma_3$ and let us pick $q=2$\footnote{The anomalous dimensions can also be obtained following \cite{Alday:2014tsa}. Both results agree.}. The above equation becomes
\begin{align}
&\kappa  \times \frac{ s t (-s-t+4)}{(s-6) (s-4) (t-6) (t-4)} =  \tilde{\gamma} (0, 2) \frac{11  \left(\frac{9 s^2+18 s (t-4)+9 t^2-72 t+160}{(s-4) (t-4)}-\frac{(s+t-5)^2}{(s-6) (t-6)}\right)}{2880 (s+t-10)}  \nonumber\\
&+\tilde{\gamma} (0, 0) \frac{ \left(\frac{7}{(s-4) (t-4)}-\frac{1}{(s-6) (t-6)}\right)}{72 (s+t-10)}   - \tilde{\gamma} (1, 0)\frac{1}{320 (s-6) (t-6) (s+t-10)} \nonumber \\
&-\tilde{\gamma} (1, 2) \frac{13  \left(11 s^2+22 s (t-5)+11 t^2-110 t+300\right)}{230400 (s-6) (t-6) (s+t-10)}\,.
\end{align}
The solution to the equation is
\begin{align}
\tilde{\gamma}(0, 0)= -\frac{256}{7} \kappa\,,\quad
\tilde{\gamma}(0,2) = \frac{640}{11} \kappa\,,\quad
\tilde{\gamma}(1,0)=-\frac{216960}{77} \kappa\,,\quad
\tilde{\gamma}(1,2)=\frac{640000}{143}  \kappa\,. 
\end{align}
A contact diagram $\kappa_{a,b}\sigma_2^a \sigma_3^b$ with $a+b < q$ on the l.h.s.\ of (\ref{MainEquation1}) is simply matched by double trace operators on the r.h.s., as we demonstrated in an example. However, if $a+b \geq q$, this is not the case. This is a crucial point, which justifies why the dispersive sum rule (\ref{MainEquation1}) \underline{must} be sensitive to operators in the UV in order to be satisfied.

If $a+b \geq q$, then double trace operators still gain anomalous dimensions due to the contact diagram, and these are easily calculable. In that case, the contact diagram on the l.h.s.\ of (\ref{MainEquation1}) is matched by the anomalous dimensions of double trace operators on the r.h.s.\ of (\ref{MainEquation1}) \underline{plus} contributions from `stringy' operators, which we examine in the next section.

Let us work this out in an example. Consider the contact diagram $\kappa \times \sigma_2$. Using a dispersion relation with $q \geq 2$, we can calculate its anomalous dimensions. We find 
\begin{align}
\tilde{\gamma}(0, 0) = -64 \kappa\,, \qquad \tilde{\gamma} (0, 2) = -\frac{320}{11}  \kappa\,.
\end{align}
We now feed these anomalous dimensions into the sum rule (\ref{MainEquation1}) at $q=1$. Only $\ell=0, 2$ and $n=m=0$ are nonzero. We find that
\begin{align}
\frac{\kappa (s^2 + t^2 + u^2)}{(s-4)(t-4)} - \sum_{\ell=0, 2}^{2} C_{0, \ell}^2  \frac{d}{d \tau} \tilde{\omega}_{\tau=4, \ell}(s, t; q; 0) \times \tilde{\gamma}(0, \ell) = -2 \kappa\,.
\end{align}
We conclude that the operators with very large twist must produce a $-2 \kappa$. We explain the precise way in which this happens in the next section.

Let us now consider all contact diagrams that appear in \eqref{M_strong_coupling} at the same time. We can obtain sum rules of increasing complexity by fixing the order in $\frac1{\lambda}$ we are considering, relative to $q$. Considering \eqref{M_strong_coupling} at $O(\lambda^{-\frac32-q})$, only the contact terms $\sigma_2^a \sigma_3^b$ with $a+b\geq q$ contribute nontrivially to the sum rule, but at the same time the flat space limit ensures that only terms with $2a+3b\leq 2q$ are possible.
So the only possible contact term is $\sigma_2^q$. At $O(\lambda^{-2-q})$ there is also $\sigma_2^{q-1}\sigma_3$, and so on. The sum rule for the first few orders in $\frac{1}{\lambda}$ reads
\bea
{}&\frac{M(s, t)}{ \prod_{i=1}^{q} (s- 2 -2i) (t - 2 -2i)} - \sum_{\text{double traces}} C_{\O_{\tau, \ell}}^2 \omega_{\tau, \ell}(s, t; q)  \\
&= \lambda^{-q-\frac{3}{2}} (-2)^q \tl\alpha_{q, 0} 
+ \lambda^{-q-2} \Big( (-2)^{q-1}  \tl\alpha_{q-1,1} \times u + (-2)^q  \tl\beta_{q, 0}  \Big) + O(\lambda^{-q-\frac{5}{2}})\,.
\eea{MELLIN MINUS DT}
We show the next two orders in $\frac{1}{\lambda}$ in appendix \ref{app:mellin_minus_dt}.

\subsection{Stringy operators}
\label{sec:stringy_operators}

The operators with low twist that are exchanged in the correlator (\ref{2222}) are the short multiplets and the double trace operators. As we have seen, these are not enough to satisfy (\ref{MainEquation1}). So, in (\ref{MainEquation1}) we need to include the contribution of operators with very large twist. One should think of these operators as the UV completion of the effective field theory made up of the low twist operators. 

We work out how to evaluate the r.h.s.\ of (\ref{MainEquation1}) when the twist $\tau$ is very large in appendix \ref{app:LargeTwistSums}. The leading term is
\begin{align}\label{SumRuleAtLargeTwist}
\lim_{\tau \rightarrow \infty}\omega_{\tau, \ell}(s, t; q) = \pi^{-3}(-1)^q 4^{\tau+ \ell+q+9}\Gamma(6+2q)  (\ell+1)  \tau^{-4(3+q)}  \sin^2\left(\frac{\pi \tau}{2}\right)\,.
\end{align}
The subleading terms in $\frac{1}{\tau}$ are easily calculable using the methods of appendix \ref{app:LargeTwistSums}.

We know from \cite{Gubser:1998bc}  that the twists of the stringy operators scale as $\tau = \tau_0 \times \lambda^{\frac{1}{4}} + ...$, where $...$ denotes subleading terms in $\lambda$ and $\tau_0$ does not depend on $\lambda$. We can use this to fix the leading $\lambda$ dependence of the OPE coefficient $\mathcal{C}^2$ of the stringy operators. Indeed, by comparing with the leading $\lambda$ term in (\ref{MELLIN MINUS DT}),
\begin{align}
\lambda^{-\frac{3}{2} - q} \sim  \mathcal{C}^2  \lambda^{-3-q}  2^{2 \tau_0 \lambda^{\frac{1}{4}}}  \sin^2\left(\frac{\pi \tau_0 \lambda^{\frac{1}{4}}}{2}\right)  \qquad \Leftrightarrow \qquad \mathcal{C}^2 \sim \frac{2^{-2 \tau_0 \lambda^{\frac{1}{4}}} \lambda^{\frac{3}{2}}}{\sin^2(\frac{\pi \tau_0 \lambda^{\frac{1}{4}}}{2})} \,.
\end{align}
The pole at $\tau_0 \lambda^{\frac{1}{4}} = 2 n$ is related to mixing between the stringy operators and double trace operators with $n >> 1$. Besides the t'Hooft coupling $\lambda$, the CFT data of the stringy operators depends on other quantum numbers, like the spin $\ell$. We will denote these collectively by the letter $r$. From the above discussion, we conclude that we can write the twists $\tau(r; \lambda)$ and the OPE coefficients $\mathcal{C}^2(r; \lambda)$ of the stringy operators as
\begin{align}
\tau(r; \lambda) &= \tau_0(r) \lambda^{\frac{1}{4}} + \tau_1(r)  +  \tau_2(r) \lambda^{-\frac{1}{4}} + \ldots \,, \label{twistsStringy}\\
\mathcal{C}^2(r; \lambda) &=  \frac{\pi ^3}{2^{12}} \frac{ 2^{-2 \tau(r; \lambda)} \tau(r; \lambda)^6 }{\sin^2(\frac{\pi \tau(r; \lambda)}{2}) } \frac{1}{2^{2 \ell}(\ell+1)}  f(r; \lambda)\,, \label{OPEStringy1} \\
 f(r; \lambda) &= f_0(r) +  f_1(r) \lambda^{-\frac{1}{4}} + f_2(r) \lambda^{-\frac{1}{2}} + \ldots \,. \label{OPEStringy2}
\end{align}
Thus, the contribution of the stringy operators to the r.h.s.\ of (\ref{MainEquation1}) is
\begin{align}\label{Stringy}
\sum_{\text{stringy} } \mathcal{C}^2(r; \lambda)  \omega_{\tau(r; \lambda), \ell}(s, t; q) =  \frac{\hat{\alpha}_{q, 0}}{\lambda^{q+\frac{3}{2}}} 
+ \frac{\hat{\chi}_q}{\lambda^{q+\frac{7}{4}}}  + \frac{ \hat{\alpha}_{q-1, 1} u + \hat{\beta}_{q,0}}{ \lambda^{q+2}}  + \ldots ,
\end{align}
where $\ldots$ denotes subleading terms in $\lambda^{-\frac{1}{4}}$\footnote{The subleading terms are fully calculable, but their expressions very lengthy.}. The coefficients $\hat{\alpha}_{q, 0}$, $\hat{\chi}_{q}$, $\hat{\alpha}_{q-1, 1}$ and $\hat{\beta}_{q,0}$ can be expressed in terms of $f_0(r)$, $f_1(r)$, $\ldots$ and $\tau_0(r)$, $\tau_1(r)$, $\ldots$. For example,
\bea
\hat{\alpha}_{q, 0} &= \sum_r (-1)^q 4^{q+3} f_0(r) \Gamma (2 q+6) \tau^{-6-4q}_0(r), \\
\hat{\chi}_{q} &= \sum_r \hat{\alpha}_{q, 0} \left( \frac{f_1(r)}{f_0(r)} -\frac{4 (2 q+3) \tau_1(r)+8 \ell (q+3)+16 q+47}{2 \tau_0(r)}  \right).
\eea{alphahat_chihat}
(\ref{Stringy}) $=$ (\ref{MELLIN MINUS DT}) gives rise to nontrivial equations like
\bea
(-2)^q \tl {\alpha}_{q, 0} &= \hat{\alpha}_{q, 0}\,, \\
0 &=  \hat{\chi}_q \,, \\
(-2)^{q-1}\tl {\alpha}_{q-1, 1} &= \hat{\alpha}_{q-1, 1}\,,\\
 (-2)^q \tl {\beta}_{q, 0} &= \hat{\beta}_{q, 0}\,.
\eea{SimplestEquation}
For instance the first equation, in terms of the rescaled coefficients $\alpha_{q,0}$, reads
\begin{equation}
\alpha_{q,0} = \sum_r \frac{4^{3+2q} f_0(r)}{\tau_0^{6+4q}(r)}\,.
\end{equation}
We can already draw some conclusions from this simple equation. First note that, since $f_0(r) \geq 0$ (since it is the rescaled squared OPE coefficient), this implies all $\alpha_{q,0}$ are positive. Furthermore, note that the large $q$ behaviour of $\alpha_{q,0}$ is controlled by the heavy operators with the minimum twist $\tau_{min,0}$. 
\begin{equation}
\alpha_{q,0} = \left(\frac{4}{\tau_{min,0}^2} \right)^{3+2q} \sum_{r_\text{min}} f_0(r) + {\cal O}\left(\left(\frac{4}{\tau_{next,0}^2} \right)^{3+2q}  \right)\,,
\end{equation}
where the sum runs over all $r$ such that $\tau_0(r)= \tau_{min,0}$. In the following section we will analyse this and other equations, with the input from flat space. 

\section{Flat space limit constraints}

\subsection{Quantisation of the twists}

In the preceding section we wrote down equations that relate the conformal data of the stringy operators to the coefficients in the strong coupling expansion of the Mellin amplitude (\ref{M_strong_coupling}). 
In this section, we analyse the simplest of these equations. At each order in $\frac{1}{\lambda}$, we consider the coefficients $\alpha_{a, b}$ multiplying the polynomials of highest degree in (\ref{M_strong_coupling}). Such coefficients are known from the flat space limit formula (\ref{flatspace_amplitude}). Our equations relate $\alpha_{a, b}$ to $f_0(r)$ and $\tau_0(r)$.
The simplest equation is the first line of (\ref{SimplestEquation}), which can be written as
\begin{align}\label{BeautifulEquation}
\zeta(3+2 a) = \sum_r \frac{4^{3+2a} f_0(r)}{\tau^{6+4a}_0(r)}\,,
\end{align}
where the sum runs over the quantum numbers $r$ that characterise the stringy operators, like the spin, and we have used the explicit form of $\alpha_{a,0}$. It is very instructive to consider the limit  $a \rightarrow \infty$. In this limit $\lim_{a \rightarrow \infty} \zeta(3+2 a) = 1$. Furthermore, on the r.h.s, we know that $f_0(r), \tau_0 (r) > 0$ due to unitarity. Let $\tau_{\text{min}, 0}$ denote the minimum value of $\tau(r)$ among all its possible values. From $\lim_{a \rightarrow \infty}$ we conclude that
\begin{align}
1 = \lim_{a \rightarrow \infty} \left(\frac{4}{\tau^2_{\text{min}, 0}}\right)^{3+2a} \sum_{r_\text{min}} f_0(r)\,,
\end{align}
where we sum over all the allowed values of $r$ such that $\tau(r) = \tau_{\text{min}, 0}$. We immediately conclude that
\begin{align}
\tau_{\text{min}, 0} =2, ~~~~~ \sum_{r_\text{min}} f_0(r) = 1 \,.
\end{align}
We can push this idea further, and expand both sides of (\ref{BeautifulEquation}) for large $a$. Since
\begin{align}
\zeta(3 + 2 a) ~ = ~ \frac{1}{1^{3 + 2 a}} ~ + ~ \frac{1}{2^{3 + 2 a}} ~ + ~ \frac{1}{3^{3 + 2 a}} ~ + ~ \ldots \,,
\end{align}
we conclude that the only possible values of $\tau_0(r)$ are
\begin{align}
\boxed{ \tau_0(r) = 2\sqrt{\delta}, ~~ \delta \in \mathbb{N} }\,.
\end{align}
This precisely agrees with the celebrated result of \cite{Gubser:1998bc} for the dimension of short strings. Note that at large $\lambda$ these strings probe a near-flat region of $AdS$, and these dimensions are found by an expansion around flat space. It is interesting to note that also in our context this set of dimensions follows from flat space physics. In summary, we learn that at leading order in $\lambda$, the stringy operators are characterised by two quantum numbers: $\delta$, which controls their twist, and $ \ell$ which controls the spin. 
 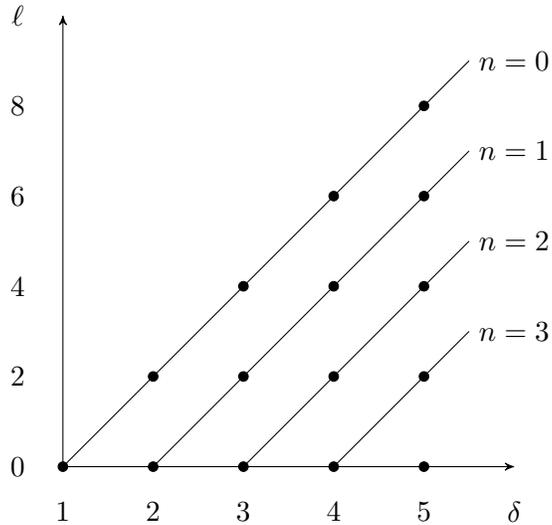
\begin{figure}
\centering
  \begin{tikzpicture}[scale=0.6]
    \coordinate (nw) at (0,10);
    \coordinate (sw) at (0,0);
    \coordinate (se) at (10,0);
    \draw[->] (sw) --  (nw) ;
    \draw[->] (sw) --  (se) ;
    \node at (0,-1) [] {$1$}; 
    \node at (2,-1) [] {$2$}; 
    \node at (4,-1) [] {$3$}; 
    \node at (6,-1) [] {$4$}; 
    \node at (8,-1) [] {$5$}; 
    \node at (10,-1) [] {$\delta$}; 
    \node at (-1,0) [] {$0$};
    \node at (-1,2) [] {$2$};
    \node at (-1,4) [] {$4$};
    \node at (-1,6) [] {$6$};
    \node at (-1,8) [] {$8$};
    \node at (-1,10) [] {$\ell$};
    \draw[-] (0,0) --  (9,9) ;
    \draw[-] (2,0) --  (9,7) ;
    \draw[-] (4,0) --  (9,5) ;
    \draw[-] (6,0) --  (9,3) ;
    \node at (10,9) [] {$n=0$};
    \node at (10,7) [] {$n=1$};
    \node at (10,5) [] {$n=2$};
    \node at (10,3) [] {$n=3$};
    \filldraw [black] (0,0) circle (3pt);
    \filldraw [black] (2,0) circle (3pt);
    \filldraw [black] (2,2) circle (3pt);
    \filldraw [black] (4,0) circle (3pt);
    \filldraw [black] (4,2) circle (3pt);
    \filldraw [black] (4,4) circle (3pt);
    \filldraw [black] (6,0) circle (3pt);
    \filldraw [black] (6,2) circle (3pt);
    \filldraw [black] (6,4) circle (3pt);
    \filldraw [black] (6,6) circle (3pt);
    \filldraw [black] (8,0) circle (3pt);
    \filldraw [black] (8,2) circle (3pt);
    \filldraw [black] (8,4) circle (3pt);
    \filldraw [black] (8,6) circle (3pt);
    \filldraw [black] (8,8) circle (3pt);
  \end{tikzpicture}
\caption{Chew-Frautschi plot of the stringy operators.} \label{fig:chew_frautschi}
\end{figure}
These stringy operators have also been extensively studied in the integrability literature, as dual to the Konishi operators and its generalisations, see for instance \cite{Gromov:2011de,Basso:2011rs,Gromov:2011bz}. Instead of $(\delta,\ell)$ it is customary to use another set of quantum numbers $(n,\ell)$ such that 
\beq
\Delta = 2 \sqrt{\delta}\, \lambda^{\frac{1}{4}} + O(\lambda^0)
= 2 \sqrt{\frac{\ell}{2} + n + 1} \, \lambda^{\frac{1}{4}} + O(\lambda^0)\,,
\eeq
where $n=0,1,2,\cdots$ labels the Regge trajectories as illustrated in figure \ref{fig:chew_frautschi}.
The leading Regge trajectory $n=0$ has been particularly well studied. In this case the dimensions are given by 
\beq
\Delta_{n=0} =  \sqrt{2\ell + 4} \lambda^{\frac{1}{4}} - 2
+ \frac{3 \ell^2 + 10 \ell + 16}{4 \sqrt{2\ell + 4}} \lambda^{-\frac{1}{4}}
+O\left(\lambda^{-\frac{3}{4}}\right)\,.
\label{integrability}
\eeq
Before proceeding, let's make the following remark. The relation $\delta=\frac{\ell}{2}+n+1$, with $n=0,1,2,\cdots$, in particular implies that, for a given $\delta$, the spin runs over $\ell=0,2,\cdots,2\delta-2$. This precise range of spins for given $\delta$ also follows from our tower of equations, namely the equations imply $f(\delta,\ell)=0$ for $\ell$ outside this range. 

\subsection{Leading OPE coefficients}
We turn to the study of the OPE coefficients at leading order in $\lambda$. Having found the allowed values for the twists, we can rewrite  (\ref{BeautifulEquation}) as
\begin{align}
\zeta(3 + 2 a) = \sum_{\delta=1}^{\infty} \frac{ \sum_{\ell=0, 2}^{2 \delta -2} f_0(\delta, \ell) }{\delta^{3+2a}}\,.
\end{align}
By matching the $a$ dependence on both sides we conclude that
\begin{align}
\boxed{ \sum_{\ell=0,2,\cdots}^{2 \delta -2} f_0(\delta, \ell) = 1, ~~ \forall \delta \in \mathbb{N} }\,.
\end{align}
This in particular implies $f_0(1, 0)=1$ but does not fully fix the OPE coefficients for higher values of $\delta$. In order to achieve that we consider the equations for  $\alpha_{a,b}$, for $b=0,1,2,\cdots$ obtained from the dispersion relations
\bea
\alpha_{a,0} ={}& \sum\limits_{\de = 1}^\infty 
\frac{1}{\delta^{3+2a}} F_0^{(0)}(\delta)\,,\\
\alpha_{a,1} ={}& \sum\limits_{\de = 1}^\infty 
\frac{1}{\delta^{6+2a}} \left(
(a+\tfrac32) F_0^{(0)}(\delta)-\frac{2}{3} F_1^{(0)}(\delta)\right)\,,\\
\alpha_{a,2} ={}& \sum\limits_{\de = 1}^\infty 
\frac{1}{\delta^{9+2a}} \left(
\frac{1}{2} (a+1) (a+3) F_0^{(0)}(\delta)-\frac{2}{3} (a+\tfrac52) F_1^{(0)}(\delta)+\frac{2}{15}F_2^{(0)}(\delta)\right)\,,\\
\alpha_{a,3} ={}& \sum\limits_{\de = 1}^\infty 
\frac{1}{\delta^{12+2a}} \bigg(
\frac{1}{6} (a+1) (a+2) (a+\tfrac92)  F_0^{(0)}(\delta)
-\frac{1}{3} (a^2 +6a+10)  F_1^{(0)}(\delta)\\
&+\frac{2}{15} (a+\tfrac72)  F_2^{(0)}(\delta)-\frac{4}{315} 
F_3^{(0)}(\delta)
\bigg)\,,
\eea{alpha_equations}
and so on. These relations were derived as explained in section \ref{sec:stringy_operators}, using the large twist expansion of Mack polynomials of appendix \ref{app:LargeTwistSums}. They depend on the OPE data through the combinations
\beq
F_m^{(0)}(\delta) = \sum_{\ell=0,2,\cdots}^{2(\de-1)}  (\ell-m+1)_m (\ell+2)_m  f_0(\de,\ell)\,.
\eeq
Since we know all the coefficients $\alpha_{a,b}$ from the flat space limit, we want to solve for the functions $F_m^{(0)}(\delta)$. To this end we rearrange the equations
\small
\begin{align}
 \sum\limits_{\de = 1}^\infty 
\frac{F_0^{(0)}(\delta)}{\delta^{3+2a}}  ={}& \alpha_{a,0}
=\zeta(2a+3)\,,\nonumber\\
-\frac23 \sum\limits_{\de = 1}^\infty \frac{F_1^{(0)}(\delta)}{\delta^{6+2a}}  ={}&
\alpha_{a,1}- \left(a+\tfrac32\right) \alpha_{a+\frac32,0}
=-2 \zeta (2 a+5,1)\,,\nonumber\\
\frac{2}{15}\sum\limits_{\de = 1}^\infty \frac{F_2^{(0)}(\delta)}{\delta^{9+2a}}  ={}&
\alpha_{a,2} - \left(a + \tfrac52 \right)\alpha_{a+\frac32,1} + \frac12 (a+3)(a+4)\alpha_{a+3,0} \label{F0_of_alpha}\\
={}& \zeta (2 a+7,2) + 4 \zeta (2 a+7,1,1)  \,,\nonumber\\
-\frac{4}{315}\sum\limits_{\de = 1}^\infty \frac{F_3^{(0)}(\delta)}{\delta^{12+2a}}  ={}&\alpha_{a,3}-\left(a+\tfrac{7}{2}\right) \alpha_{a+\frac{3}{2},2}+\frac{1}{2} (a+3) (a+6) \alpha_{a+3,1}
-\frac{1}{6} (a+\tfrac72) (a+\tfrac92) (a+\tfrac{17}{2}) \alpha_{a+\frac{9}{2},0}\nonumber\\
={}&
-2 \zeta(2 a+9,1,2)-2\zeta(2 a+9,2,1)-8 \zeta(2 a+9,1,1,1)\,,\nonumber
\end{align}
\normalsize
where we inserted the results for the $\alpha_{a,b}$ \eqref{alphas}. Remarkably, the linear combinations of $\alpha_{a,b}$ emerging from the equations are such that each sum over $F_m^{(0)}(\delta)$ is given by a sum of multiple zeta values with uniform first argument $\zeta(2a+2m+3,\cdots)$. By comparing \eqref{F0_of_alpha} with the definition \eqref{MZV} it is also apparent that $F_m^{(0)}(\delta)$ are given by nested sums. Indeed, defining the nested sums
\beq
\label{nested}
Z_{s_1,s_2,s_3,\ldots} (N) = \sum\limits_{n=1}^{N} \frac{Z_{s_2,s_3,\ldots} (n-1)}{n^{s_1}} \,, \qquad Z (N) = 1\,, \qquad Z_{s_1,s_2,\ldots} (0)=0\,,
\eeq
which naturally lead to multiple zeta values
\beq
\zeta(s,s_1,s_2,\ldots) = \sum\limits_{\delta=1}^\oo \frac{Z_{s_1,s_2,\ldots} (\delta-1)}{\delta^{s}} \,,
\label{Z_to_zeta}
\eeq
we see that one can define 
\beq
F_m^{(0)}(\delta) = \frac{2^{m+1} \Gamma(m+\frac{3}{2})}{\sqrt{\pi}} \delta^m W_m(\delta) \,,
\eeq
such that $W_m(\delta)$ are given by linear combinations of nested sums $Z_{s_1,s_2,\ldots} (\delta-1)$ of total weight $m$ and coefficients $s_i=1,2$. More precisely we have chosen our normalisations such that 
\begin{equation}
W_0(\delta)=1\,,~~~~~W_1(\delta) =Z_{1} (\delta-1) \,,
\end{equation}
and higher $W_m(\delta)$ are given by the recursion relation
\beq
W_m(\delta+1) - W_m(\delta) = \frac{m}{\delta} W_{m-1}(\delta)+
\frac{m(m-1)}{4\delta^2} W_{m-2}(\delta)\,,
\eeq
which determines the coefficient $c_m(\delta)$ in
\beq
W_m(\delta) = \sum\limits_{n=1}^{\delta-1} c_m(n)\,.
\eeq
From these recursions one can check
\bea
{}&\sum\limits_{\delta=1}^\oo \frac{W_0(\delta)}{\delta^s} = \zeta(s)\,, \quad
\sum\limits_{\delta=1}^\oo \frac{W_1(\delta)}{\delta^s} = \zeta(s,1)\,, \quad
\sum\limits_{\delta=1}^\oo \frac{W_2(\delta)}{\delta^s} = 2\zeta(s,1,1)+\frac12 \zeta(s,2)\,, \quad\\
&\sum\limits_{\delta=1}^\oo \frac{W_3(\delta)}{\delta^s} = 6\zeta(s,1,1,1)+\frac32 \zeta(s,1,2)+\frac32 \zeta(s,2,1)\,,
\eea{W_sums}
and so on, from which we can explicitly check the equations \eqref{F0_of_alpha}. 

Concerning the individual OPE coefficients $f_0(\de,\ell)$, note that $F_m^{(0)}(\delta)$ is a sum over $\delta$ different spins, so $F_m^{(0)}(\delta)$ for $0\leq m < n$ fixes all the OPE coefficients with $\delta \leq n$. We find precise agreement with the OPE coefficients computed using the method of \cite{Costa:2012cb,Goncalves:2014ffa} which is reviewed and extended in appendix \ref{app:f0}. The appendix also contains explicit formulae for the OPE coefficients.

\subsection{$f_1(\delta, \ell)$ and $\tau_1(\delta, \ell)$}

Note that (\ref{MELLIN MINUS DT}) vanishes identically at orders $\lambda^{-\frac{7}{4}-q - \frac{n}{2}}$, $n \in \mathbb{N}_0$, while (\ref{Stringy}) does not. This gives rise to non trivial equations, which turn out to be simple to solve. For example, from the order $\lambda^{-\frac{7}{4}-q}$ we obtain the equations
\begin{align}
\sum_{\ell=0, 2}^{2 \delta -2} f_0(\delta, \ell) \Big(2 + \ell + \tau_1(\delta, \ell) \Big) =0 \,, ~~ \sum_{\ell=0, 2}^{2 \delta -2} \Big( f_0(\delta, \ell) (23 + 12 \ell) -4 \sqrt{\delta} f_1(\delta, \ell) \Big) = 0\,.
\end{align}
valid for $\delta =1, 2, \ldots$. At $\delta=1$, we can immediately fix 
\begin{align}
f_1(\delta=1, \ell=0) = \frac{23}{4}\,, ~~~~ \tau_1(\delta=1, \ell =0) = -2\,.
\end{align}
We studied the tower of equations of this kind and determined $f_1(\delta, \ell)$ and $\tau_1(\delta, \ell)$ for high values of $\delta$ and $\ell=0, 2, \ldots , 2 \delta-2$. It turns out that
\begin{align}
\tau_1(\delta, \ell) = - 2 - \ell\,, ~~~~~ f_1( \delta, \ell) = f_0(\delta, \ell) \frac{3 \ell+\frac{23}{4}}{\sqrt{\delta }}\,.
\end{align}
In drawing this conclusion we assume that the degeneracy at $\lambda^{\frac{1}{4}}$ can only be lifted at order $\lambda^{-\frac{1}{4}}$, since corrections to string states should come in powers of $\alpha'$, or equivalently $\frac{1}{\sqrt{\lambda}}$. The finite shift $\tau_1(\delta, \ell) = - 2 - \ell$ simply means that the string states are dual to descendants of the R-symmetry singlet superprimaries that we are considering. 


\section{Into AdS}

\subsection{The analytic structure of the stringy corrections}

Let us start by making the following remark about a curious analyticity property of the stringy corrections in the strong coupling expansion (\ref{M_strong_coupling}) of the Mellin amplitude. Consider a contact term $\frac{\Gamma(2a + 3b + 6)}{8^{a+b}} \frac{\xi_{a,b}}{\lambda^{\frac{3}{2}+q'}} \sigma_2^a \sigma_3^b$. The sum rules always take the form
\begin{align}\label{seriesXi}
\xi_{a,b} ~ = ~ \sum_{\delta=1}^{\infty} \frac{p_{\xi,b}(a,\delta) }{\delta^{r + 2 a}},
\end{align}
where $p_{\xi,b}(a,\delta)$ is a polynomial in $a$ that depends on the conformal data of the stringy operators and $r$ is some integer, chosen such that $p_{\xi,b}(a,\delta)$ grows at most logarithmically for large $\delta$. The series on the r.h.s.\ converges for $a > \frac{1}{2} - \frac{r}{2}$ and actually naturally defines an analytic function in this region.\footnote{By contrast, we do not have any indication that $\xi_{a,b}$ should be viewed as an analytic function of $b$. } It is instructive to consider the case where $a$ is a negative integer with $a + b \geq 0$. In that case, it turns out from the sum rules that the r.h.s.\ of (\ref{seriesXi}) still converges and the sum rules for (\ref{MainEquation1}) with $q=a+b$ imply that it actually vanishes. In equations,
\begin{align}\label{Zeros}
\xi_{a, b} =0, ~~~ \text{for} ~~~ a = -b, ~ -b +1, ~ \cdots,~ -1 .  
\end{align}
As an example consider the coefficients fixed by the flat space limit
\begin{align}
&\alpha_{a,0}=\zeta(2a+3) \,,\nonumber\\
&\alpha_{a,1}=-2\zeta(2a+5,1)+\left(a+\tfrac{3}{2} \right)\zeta(2a+6)\,, \\
&\alpha_{a,2}= \zeta (2 a+7,2)+4 \zeta (2 a+7,1,1)-(2 a+5) \zeta (2 a+8,1)+\tfrac{1}{2} (a+1) (a+3) \zeta (2 a+9)\,. \nonumber
\end{align}
They are, indeed, analytic functions of $a$, which can actually be analytically extended to the whole complex plane (except at points were the zeta functions diverge). This is the case for other examples as well. Furthermore, by using classical identities one can show $\alpha_{-1,1}=\alpha_{-2,2}=\alpha_{-1,2}=0$, in agreement with our discussion.

Before proceeding, let's make the following remark. While the constraints $\alpha_{a,b}=0$ for $a=-1,\cdots,-b$ are clear from the dispersive sum rules, they are not obvious in terms of the CFT-data. For instance, $\alpha_{-1,1}=0$ implies
\begin{equation}
 \sum\limits_{\de = 1}^\infty 
\frac{1}{\delta^{4}} \left(
\frac{1}{2}F_0^{(0)}(\delta)-\frac{2}{3} F_1^{(0)}(\delta)\right)=0,
\end{equation}
which is a priori not clear, but can be readily verified for our solutions. 

\subsection{Corrections to the Virasoro-Shapiro amplitude}

We now consider $AdS$ corrections to the flat space Virasoro-Shapiro amplitude. We obtain one sum rule for each correction term, but for simplicity we focus on the first layer of corrections $\beta_{a,b}$. For the first few cases we obtain
\begin{align}
\beta_{a,0} ={}&
\sum\limits_{\de = 1}^\infty 
\frac{1}{ \delta ^{4+2a}}  \bigg(F_0(\delta )-(2 a+3) T_0(\delta )\label{equation1}\\
&+\frac18   \left(16 a^2+44 a-139\right) \delta W_1(\delta)-\frac{1}{96}\left(128 a^3+384 a^2-416 a+639\bigg)
   W_0(\delta)\right)\,,\nonumber\\
   \beta_{a,1} ={}&
\sum\limits_{\de = 1}^\infty 
\frac{1}{192\delta ^{7+2 a}}  \bigg(32 ((6 a+9) F_0(\delta )-4 F_1(\delta )-2 (a+3) ((6 a+9) T_0(\delta )-4 T_1(\delta )))\nonumber\\
&-48  \left(16 a^2+84 a-157\right) \delta ^2 W_2(\delta)+8   \left(112 a^3+732 a^2+455 a+1215\right) \delta W_1(\delta)\nonumber\\
&-\left(256 a^4+2304 a^3+6464 a^2+9918 a+7389\right) W_0(\delta)\bigg)\,,\label{equation2}\\
\beta_{a,2} ={}&
\sum\limits_{\de = 1}^\infty
\frac{1}{30 \delta ^{10+2a}}  \Big(15 (a+1) (a+3) F_0(\delta )-10 (2 a+5) F_1(\delta )+4 F_2(\delta )\nonumber\\
&-(2 a+9) (15 (a+1) (a+3) T_0(\delta )-10 (2 a+5) T_1(\delta )+4 T_2(\delta ))\Big)\nonumber\\
&+\frac{1}{192 \delta ^{10+2a}}  \bigg(48  \left(16 a^2+124 a-127\right) \delta ^3 W_3(\delta)\nonumber\\
&-4   \left(320 a^3+3216 a^2+5716 a+6189\right)
  \delta^2 W_2(\delta)\nonumber\\
   &+2 \left(352 a^4+4552 a^3+18518 a^2+36614 a+35409\right) \delta W_1(\delta)\nonumber\\
   &-\left(128 a^5+2048 a^4+12256 a^3+36319 a^2+52956 a+26973\right) W_0(\delta)\bigg)\,, \label{equation3}
\end{align}
and so on. We have defined the following sums involving OPE data
\bea
T_m(\delta ) ={}& \sum_{\ell=0,2}^{2(\de-1)} \sqrt{\de}  (\ell-m+1)_m (\ell+2)_m  f_0(\de,\ell) \tau_2(\de,\ell)\,,\\
\qquad F_m(\delta ) ={}& \sum_{\ell=0,2}^{2(\de-1)}\de  (\ell-m+1)_m (\ell+2)_m  f_2(\de,\ell)\\
& - \frac{39}{4} \ell \left( \delta_{m,0}+ \delta_{m,1} (\ell^2-4)+ \delta_{m,2} (\ell^2-4) (\ell^2-11) + \ldots\right) f_0(\de,\ell)\,.
\eea{T2F2}
Although the system (\ref{equation1}-\ref{equation3}) has been derived for integer values of $a$, note that the r.h.s.\ of these equations admits an analytic continuation in $a$, as discussed above. This allows us to consider linear combinations of the above equations to obtain simpler equations that only depend on OPE data in terms of the combination $F_m(\delta) - (2a+3m+3) T_m(\delta)$
\begin{align}
{}&\beta_{a,0} ={}
\sum\limits_{\de = 1}^\infty 
\frac{1}{ \delta ^{4+2a}}  \bigg(F_0(\delta )-(2 a+3) T_0(\delta )\label{beta0_subtracted}\\
&+\frac18   \left(16 a^2+44 a-139\right) \delta W_1(\delta)-\frac{1}{96}\left(128 a^3+384 a^2-416 a+639\bigg)
   W_0(\delta)\right)\,,\nonumber\\
{}&\beta_{a,1} - \left(a+ \tfrac32\right) \beta_{a+\frac32,0}
= \sum\limits_{\de = 1}^\infty \frac{1}{\delta ^{7+2a}} \bigg(  -\frac{2}{3}  (F_1(\delta ) -(2a+6) T_1(\delta ))- \left(4 a^2+21 a-\tfrac{157}{4}\right) \delta^2 W_2(\delta)\nonumber\\
&+\frac{1}{48} \left(128 a^3+768 a^2+304 a+2763\right) \delta W_1(\delta)-\left(2 a^2+13 a+18\right) W_0(\delta) \bigg)\,,
\label{beta1_subtracted}\\
&\beta_{a,2}-\left(a+\tfrac{5}{2}\right) \beta_{a+\frac{3}{2},1}+\frac{1}{2} (a+3)(a+4) \beta_{a+3,0} =\sum\limits_{\de = 1}^\infty \frac{1}{\delta ^{10+2a}} \bigg(
\frac{2}{15}  (F_2(\delta)-(2 a+9) T_2(\delta ))\nonumber\\
&+\frac{1}{4} \left(16 a^2+124 a-127\right) \delta^{3} W_3(\delta)
-\frac{1}{48} \left(128 a^3+1152 a^2+1696 a+6039\right) \delta^{2} W_2(\delta)\nonumber\\
&+\left(4 a^2+28 a+63\right) \delta W_1(\delta)
+\frac{1}{2} \left(6 a^2+57 a+126\right) W_0(\delta)
   \bigg)\,.
\label{beta2_subtracted}
\end{align}
and so on. We denote the combinations appearing on the l.h.s.\ of these equations by $\bar \beta_{a,b}$. These are the same combinations that appeared in the previous section, for the coefficients $\alpha_{a,b}$. Note that the r.h.s.\ of these subtracted equations is quadratic in $a$, except for a single cubic term. Equations for the higher coefficients $\beta_{a,b}$ follow the same pattern. Several comments are in order. First, these equations put strong constraints on the form of $\beta_{a,b}$. Indeed, in a large $a$ expansion the combinations $\bar \beta_{a,b}$ behave as
\begin{equation}
\bar \beta_{a,b} = \sum_{\delta=1}^\infty \frac{a^3 h_3(\delta)+a^2 h_2(\delta)+a h_1(\delta)+ h_0(\delta)}{\delta^{2a+3b+4}}\,,
\end{equation}
 with $h_3(\delta),h_2(\delta)$ completely determined by the equations (\ref{beta2_subtracted}). Second, note that  $h_1(\delta),h_0(\delta)$, and hence the full $\beta_{a,b} $, are given in terms of the combinations $F_m(\delta) - (2a+3m+3) T_m(\delta)$, which follow, in principle, from results of integrability. They are given by the dimensions and OPE coefficients of the heavy stringy operators to order $1/\lambda^{1/4}$. Third, note that the constraints $\beta_{-1,1}=0$ and so on, impose some constrains on the CFT data, which appear to be non-trivial from the integrability perspective. Finally, the r.h.s.\ of the equations can be written as a contribution from the combinations $F_m(\delta) - (2a+3m+3) T_m(\delta)$ plus a contribution explicitly given in terms of multiple zeta values, of uniform transcendentality. This is a common feature of observables in ${\cal N}=4$ SYM and related to its modularity properties.
Based on this it is reasonable to assume that $\beta_{a,b}$ is given as a combination of multiple zeta values of weight $4+2a+3b$ and maximal depth $b+2$. This is compatible with the known value $\beta_{1,0}=-2 \zeta(3)^2$ which has weight six and depth two.

Inspired by the expressions for $\alpha_{a,b}$ in \eqref{alphas},
let us also make the assumption that $\beta_{a,0}$ can be written in terms of sums over products of zeta values of odd arguments and make the ansatz
\beq
\beta_{a,0} = (\kappa_0 + \kappa_1 a + \kappa_2 a^2)\sum\limits_{\substack{i_1,i_2=0\\i_1+i_2=a-1}}^{a-1}\zeta(3+2i_1) \zeta(3+2i_2)
+ \kappa_3 \sum\limits_{\substack{i_1,i_2=0\\i_1+i_2=a-1}}^{a-1} i_1 i_2 \zeta(3+2i_1) \zeta(3+2i_2)\,.
\label{beta_a0_ansatz}
\eeq
We can use the expansions
\begin{align}
\sum\limits_{\substack{i_1,i_2=0\\i_1+i_2=a-1}}^{a-1}\zeta(3+2i_1) \zeta(3+2i_2)
=& \sum\limits_{\de = 1}^\infty \frac{1}{2 \delta^{4+2a}}  \left(2 a+1-4 \delta  Z_1(\delta -1)\right)\,,\label{zeta_expansions}\\
\sum\limits_{\substack{i_1,i_2=0\\i_1+i_2=a-1}}^{a-1} i_1 i_2 \zeta(3+2i_1) \zeta(3+2i_2)
=& \sum\limits_{\de = 1}^\infty \frac{1}{24 \delta^{4+2a}}  \Big(4 (a-1)^3-a+25+48 a \delta  Z_1(\delta -1)\nonumber\\
&+12 (2 a+1) \delta ^2 Z_2(\delta -1)+24 \delta ^3 Z_3(\delta -1)-24 \delta ^3 \zeta (3)\Big)\,,\nonumber
\end{align}
and match the $a^3$, $a^2$ and $a^2 \delta Z_1(\delta -1)$ terms between \eqref{equation1} and \eqref{beta_a0_ansatz} to fix three coefficients. A further one can be fixed by imposing $T_0(1) = 2$ from integrability \eqref{integrability}. We find
\beq
\kappa_0 = \frac72\,, \quad
\kappa_1 = -\frac92\,, \quad
\kappa_2 = -1\,, \quad
\kappa_3 = -2\,.
\eeq
We can now read off
\bea
T_0(\delta ) &= 2+\frac{\delta  Z_1(\delta -1)}{4}+\delta ^2 Z_2(\delta -1)\,,\\
F_0(\delta ) &= \frac{405}{32}+\frac{89 \delta  Z_1(\delta -1)}{8}+2 \delta ^2 Z_2(\delta -1)-2 \delta ^3 Z_3(\delta -1)+2 \delta ^3 \zeta (3)\,.
\eea{OPE_sums_0}
Plugging this back into \eqref{equation1} and doing the sum over $\delta$ using \eqref{Z_to_zeta} one can also write $\beta_{a,0}$ as
\bea
\beta_{a,0}={}&2 \zeta (2 a+1)\zeta (3) -2 \zeta (2 a+1,3)-(2 a+1) \zeta (2 a+2,2)\\
&+(a-1) (2 a+7) \zeta (2 a+3,1)-\frac{1}{3} a \left(4 a^2+12 a-1\right) \zeta (2 a+4)\,.
\eea{beta_a0_result}
As a non-trivial check we find the values known from localisation
\beq
\beta_{0,0} = 0\,, \qquad \beta_{1,0} = -2\zeta(3)^2\,.
\eeq
The result \eqref{OPE_sums_0} implies $f_2(1,0)=\frac{405}{32} + 2 \zeta(3)$ for the correction to the OPE coefficient of the Konishi operator.

Next we consider the equation \eqref{equation2} for $\beta_{a,1}$
where, using \eqref{OPE_sums_0}, the only unknowns are now $T_1(\delta )$ and $F_1(\delta )$.
The ansatz analoguous to \eqref{beta_a0_ansatz} is of the form
\bea
\beta_{a,1} ={}&
(\mu_0+\mu_1 a + \mu_2 a^2)
\sum\limits_{\substack{i_1,i_2,i_3=0\\i_1+i_2+i_3=a-1}}^{a-1}\zeta(3+2i_1) \zeta(3+2i_2) \zeta(3+2i_3)\\
&+\mu_3 \sum\limits_{\substack{i_1,i_2,i_3=0\\i_1+i_2+i_3=a-1}}^{a-1} (i_1^2+i_2^2+i_3^2)\zeta(3+2i_1) \zeta(3+2i_2) \zeta(3+2i_3)\\
&+(\mu_4+\mu_5 a + \mu_6 a^2 + \mu_7 a^3 + \mu_8 a^4) \zeta(7+2a)\\
&+ \text{ terms with overall factor } \zeta(3),\zeta(5), \ldots\,.
\eea{beta_a1_ansatz}
This ansatz can not be matched to the known $a^2, a^3$ and $a^4$ terms in \eqref{equation2}. Hence our assumptions are too strong and $\beta_{a,1}$ cannot be written in terms of single zeta values of odd arguments with overall weight $7+2a$ and maximal depth $3$.

Another approach would be to make an ansatz for $T_m(\delta )$ and $F_m(\delta )$ in terms of the nested sums \eqref{nested}, which would also imply that $\beta_{a,b}$ has a representation in terms of multiple zeta values. However, the constraints \eqref{Zeros} are generally not enough to fix all coefficients in such an ansatz.
We expect that it is the assumption that $\beta_{a,b}$ has a representation in terms of single zeta values that breaks down for $b>0$.

\section{Conclusions}

In this paper we have derived an infinite set of sum rules in Mellin space for the four-point function of the stress tensor in ${\cal N}=4$ SYM, to leading order in the inverse power of the central charge, but including all order corrections in $1/\lambda$. These sum rules give the explicit connection between the coefficients of the  AdS Virasoro-Shapiro amplitude in a $1/\lambda$ expansion and the CFT data for heavy string states, of dimension $\Delta \sim \lambda^{1/4}$, given in principle by integrability. The structure of the equations is such that, regardless of this data, it imposes strong constraints on the coefficients of the AdS Virasoro-Shapiro amplitude. In the same way, it imposes non-trivial constraints on the CFT data entering the equations, which appear to be non-trivial from the point of view of integrability. An interesting aspect of these relations is that the heavy operator with the lowest dimension, namely the Konishi operator at strong coupling, controls the coefficient of $\sigma_2^a \sigma_3^b$ at large $a$. 

The part of the AdS Virasoro Shapiro amplitude fixed by the flat space limit, namely the coefficients denoted $\alpha_{a,b}$, is directly related to the leading order dimensions and OPE coefficients of the heavy operators, and our relations completely determine them. In particular, the OPE coefficients $\langle {\cal O}_2 {\cal O}_2 {\cal O}^{heavy}_{\delta,\ell} \rangle$ at leading order can be seen to satisfy beautiful relations involving harmonic sums. It would be interesting to understand these relations from a different perspective. Going into $AdS$, we have considered in detail the first layer of corrections to the flat space Virasoro-Shapiro amplitude. This layer is related to the first non-trivial $1/\lambda$ corrections to the CFT data mentioned above. We have constructed solutions consistent with the results from integrability, localisation and the transcendental structure of the equations, which agrees with the expected transcendental structure of the CFT-data. For the coefficients $\beta_{a,0}$ we have found a unique solution (under these assumptions), while higher $\beta_{a,b}$ are also highly constrained in their form and could be fixed when more data becomes available from integrability.  

The application of dispersive sum rules to ${\cal N}=4$ SYM at strong coupling gives a handle on the so far elusive AdS Virasoro-Shapiro amplitude. There are many open problems that would be interesting to address. Below we list some of them in no particular order. 

Beyond the leading Regge trajectory we expect that heavy operators become degenerate at large $\lambda$. From a string theory computation we would expect this degeneracy to be lifted at order $\lambda^{-1/4}$. Understanding this lifting will be important when analysing the equations for the coefficients $\gamma_{a,b}$, the next layer of corrections to the flat space Virasoro-Shapiro amplitude. A very specific way to do this, which would be interesting on its own right, is by studying dispersive sum rules for more general correlators involving KK-modes. This should also allow to make contact with \cite{Abl:2020dbx}.

We have seen that the coefficients $\xi_{a,b}$ of the low energy expansion admit an analytic continuation in $a$, however, we used this analyticity only mildly. It would be interesting to explore this further. Relatedly, the appearance of multiple zeta values in our relations seems to be a manifestation of the rich modular properties of amplitudes in type IIB string theory and ${\cal N}=4$ SYM correlators, see \cite{Dorigoni:2022iem}.  It would be interesting to make this connection more clear. In the same direction, it would be interesting to relate our approach to results for integrated correlators from localisation, see \cite{Binder:2019jwn}.

It would be interesting to study non-perturbative corrections to the amplitude considered in this paper. In particular we expect exponentially suppressed contributions arising from the exchange of single trace operators with very large spin, whose dimension at large $\lambda$ grows like $\Delta \sim \sqrt{\lambda} \log \ell$, see  \cite{Alday:2007mf}.

It would be interesting to study, via integrability methods, single trace operators in the sub-leading Regge trajectories at strong coupling. This should correspond to length four operators in the singlet representation, of the form $Tr \left( \varphi^I\varphi^I\varphi^J\varphi^J \right)$ and $Tr \left( \varphi^I\varphi^J\varphi^I\varphi^J \right)$ and their generalisations with higher spin. 

A related system where one could derive analogous sum rules is the ABJM theory \cite{Aharony:2008ug}, in the regime where it is holographically dual to type IIA string theory on $AdS_4 \times CP^3$. Holographic correlators in this system were studied in \cite{Binder:2019mpb} and integrability methods are very well developed, see for instance \cite{Bombardelli:2017vhk}.

\section*{Acknowledgements} 

We thank Vasco Gonçalves for useful conversations. JS also thanks Alexander Dannin. The work of LFA and TH is supported by the European Research Council (ERC) under the European Union's Horizon
2020 research and innovation programme (grant agreement No 787185). LFA is also supported in part by the STFC grant ST/T000864/1.
The work of JS is supported by an early postdoc mobility fellowship (grant P2ELP2199748) by the Swiss National Science Foundation (SNSF).

\appendix

\section{Mack polynomials}\label{app:Mack appendix}

We write out some definitions used in the main text.
\begin{align}
\mathcal{Q}_{\ell, m}^{\tau, d}(t) \equiv -\frac{2^{3 \ell+2 \tau -2} \Gamma \left(\ell+\frac{\tau }{2}-\frac{1}{2}\right) \Gamma \left(\ell+\frac{\tau }{2}+\frac{1}{2}\right) \Gamma \left(-\frac{d}{2}+\ell+\tau +1\right) Q_{\ell, m}^{\tau, d}(t)}{\pi  \Gamma (m+1) \Gamma \left(\ell+\frac{\tau }{2}\right)^2 \Gamma (\ell+\tau -1) \Gamma \left(-m-\frac{\tau }{2}+4\right)^2 \Gamma \left(-\frac{d}{2}+\ell+m+\tau +1\right)}. 
\end{align}
$Q_{\ell, m}^{\tau, d}(t)$ is called a Mack polynomial in the literature \cite{Mack:2009mi}. We found the following representation useful \cite{Dey:2017fab}
\begin{align}
&Q_{\ell, m}^{\tau, d}(t) = (-1)^{\ell} 4^{\ell} \sum _{n_1=0}^{\ell} \sum_{m_1=0}^{\ell-n_1} (-m)_{m_1} \left(m+\frac{t}{2}+\frac{\tau }{2}\right)_{n_1} \tilde{\mu}(\ell,m_1,n_1,\tau ,d), \\
&\tilde{\mu}(\ell,m, n,\tau ,d) \equiv \frac{2^{-\ell} \Gamma (\ell+1) (-1)^{m+n} \left(\ell-m+\frac{\tau }{2}\right)_m \left(n+\frac{\tau }{2}\right)_{\ell-n} }{\Gamma (m+1) \Gamma (n+1) \Gamma (\ell-m-n+1)}  \\
& \times \left(\frac{d}{2}+\ell-1\right)_{-m} (2 \ell+\tau -1)_{n-\ell} \left(m+n+\frac{\tau }{2}\right)_{\ell-m-n} \nonumber \\
& \times \, _4F_3\left(-m,-\frac{d}{2}+\frac{\tau }{2}+1,-\frac{d}{2}+\frac{\tau }{2}+1, \ell+n+\tau -1;\ell-m+\frac{\tau }{2},n+\frac{\tau }{2},-d+\tau +2;1\right) . \nonumber 
\end{align}

\section{Polyakov conditions}\label{app:Polyakov conds}

In this appendix we study Polyakov conditions for the Mellin amplitude (\ref{MellinDef}). The discussion is very similar to the one in \cite{Penedones:2019tng,Carmi:2020ekr} and it applies for any central charge and gauge coupling. Let us study the first Polyakov condition at $s=4$. The Mellin transform of the physical correlator is equal to
\begin{align}
\hat{M}(s, t) \equiv M(s,t) \times \Gamma^2\Big(2- \frac{s}{2}\Big) \Gamma^2\Big(2- \frac{t}{2}\Big) \Gamma^2\Big(2- \frac{u}{2}\Big)\,.
\end{align}
It is useful to consider $\hat{M}(s, t)$, because, as argued in \cite{Penedones:2019tng}, it is expected that it only has simple poles at the locations of the physical operators. In the limit $s \rightarrow 4$, $\hat{M}(s, t)$ has an accumulation of poles and it develops a branch point behaviour, that we will study. More precisely, we will consider $\tilde{M}(s,t) \equiv (s-4) \times \hat{M}(s,t)$, for reasons that will be clear in a moment. The quantity
\begin{align}\label{sumInfinitePoles}
\sum_{\ell = \text{finite}}^{\infty} \frac{\Res ~ \tilde{M}\big(s = 4 + \gamma(\ell) , t \big)}{s-4 - \gamma(\ell)}\,,   
\end{align}
where we sum over the residues of $\tilde{M}$, captures the infinite sum over poles. $\gamma(\ell)$ represents the anomalous dimension of the spin $\ell$ operator in the leading Regge trajectory. At large spin, $\gamma(\ell)$ is controlled by the exchange of the stress tensor. Thus, $\gamma(\ell)= - \frac{\chi}{\ell^2}$, where $\chi>0$. In the large $\ell$ limit (\ref{sumInfinitePoles}) becomes
\begin{align}
\int_{\text{finite}}^{\infty} d\ell \frac{\frac{4}{3} \chi  \ell^{-u-3} \Gamma \left(\frac{u}{2}+2\right)^2 +\frac{4}{3} \chi  \ell^{u-3} \Gamma \left(2-\frac{u}{2}\right)^2}{s-4 + \frac{\chi}{\ell^2}}\,.
\end{align}
This integral converges when $-2<Re(u)<2$. This finite width of convergence is the reason we chose to consider $\tilde{M}(s, t)$, instead of $\hat{M}(s, t)$.

We can express this integral in terms of hypergeometric functions and afterwards take the limit $s \rightarrow 4$. We find that
\begin{align}
\lim_{s \rightarrow 4} \tilde{M}(s, t) \approx (s-4)^{\frac{u}{2}} + (s-4)^{-\frac{u}{2}} + \text{reg}\,.
\end{align}
$\text{reg}$ denotes regular terms of order $(s-4)^0$ that are not captured by (\ref{sumInfinitePoles}). Since $\lim_{s \rightarrow 4} M(s, t) = \lim_{s \rightarrow 4} (s-4) \tilde{M}(s, t) $, then
\begin{align}
M(s, t) \approx (s-4)^{1+\frac{u}{2}} + (s-4)^{1-\frac{u}{2}} + \text{reg'}\,,
\end{align}
where $\text{reg'}$ are regular terms of order $(s-4)^1$. We conclude that 
\begin{align}
\lim_{s \rightarrow 4} M(s,t) = 0, ~~ -2 < \text{Re}(u) < 2\,.
\end{align}
The same applies to $\lim_{t \rightarrow 4} M(s,t)$. As explained in \cite{Carmi:2020ekr}, at $s, t= 6, 8, ...$ the strip in the $u$ plane where we can apply the Polyakov condition is actually larger than for $s, t=4$. The conclusion of this exercise is that there are no contributions to (\ref{MainEquation1}) from $s,t =4, 6, 8, ...$ provided $-2 < \text{Re}(u) < 2$.

\section{Equation (\ref{MELLIN MINUS DT})}
\label{app:mellin_minus_dt}

We write two more orders in equation (\ref{MELLIN MINUS DT})
\bea
&\frac{M(s, t)}{ \prod_{i=1}^{q} (s- 2 -2i) (t - 2 -2i)} - \sum_{\text{double traces}} C_{\O_{\tau, \ell}}^2 \omega_{\tau, \ell}(s, t; q)  \\
&= (-2)^q \lambda ^{-q-\frac{3}{2}} \tl\alpha_{q,0}
+(-2)^{q-1} \lambda ^{-q-2} \left(u \tl\alpha _{q-1,1}-2 \tl\beta _{q,0}\right)\\
&+(-2)^{q-2} \lambda ^{-q-\frac{5}{2}} \bigg(4 \tl\gamma _{q,0}+u^2 \tl\alpha _{q-2,2}-2 u \tl\beta _{q-1,1}\\
&+\frac{8}{3} \tl\alpha _{q+1,0} \left(4 q^3+3 q^2 (u+2)+q (3 (u-1) u+14)+3 \left(s^2+(s+u) (u-4)+8\right)\right)\bigg)\\
&+(-2)^{q-3} \lambda ^{-q-3} \bigg(\frac{8}{3} u \tl\alpha _{q,1} \left(4 q^3+3 q^2 (u+2)+q (3 (u-1) u+14)+3 s (s+u-4)\right)\\
&+\frac{16}{3} \tl\beta _{q+1,0} \left(-4 q^3-3 q^2 (u+2)+q (-3 (u-1) u-14)-3 \left(s^2+(s+u) (u-4)+8\right)\right)\\
&+u^3 \tl\alpha _{q-3,3}-2 u^2 \tl\beta _{q-2,2}+4 u \tl\gamma _{q-1,1}-8 \tl\delta _{q,0}\bigg) + O\left(\lambda ^{-q-\frac{7}{2}} \right)\,.
\eea{MELLIN MINUS DT2}

\section{Large twist sums}\label{app:LargeTwistSums}

The main trick to expand (\ref{defSumRule}) at large twist $\tau$ is to recognise that the sum in $m$ is dominated by $m \sim \tau^2$. Replacing $m= x \tau^2$, the relevant Mack polynomials at large twist are given by
\small
\begin{align}
Q_{\ell, m=x \tau^2}^{\tau, d=4}(u-4) \approx \tau^{2 \ell} x^{\ell} \frac{1 + \ell}{2^{\ell}} \left(1 + \frac{\ell}{2 x \tau} - \frac{\ell (1 -\ell +(6  +6 \ell  -4 u  -2 \ell u) x+ (\ell-1) x^2)}{6 x^2 \tau^2} + O\left(\frac{1}{\tau^3}\right) \right).
\end{align}
\normalsize
We can derive many more orders, but it becomes cumbersome to write them.
We then turn the sum $\sum_{m=0}^{\infty}$ in (\ref{defSumRule}) into $\int_0^{\infty} dx \tau^2$. Furthermore we use the integral formula 
\begin{align}
\int_0^{\infty} dx x^a e^{- \frac{x}{4}} = 4^{-1-a} \Gamma(-1-a)\,. 
\end{align}
In this manner we can derive the leading and many subleading orders in (\ref{SumRuleAtLargeTwist}).

\section{OPE coefficients from flat space limit of conformal partial waves}
 \label{app:f0}

In \cite{Costa:2012cb,Goncalves:2014ffa} the OPE coefficients of single trace operators at large $\lambda$ in the correlator of four Lagrangians were computed by matching the flat space limit of the conformal partial wave expansion to the partial wave expansion of the flat space scattering amplitude. In this appendix we apply the method directly to the reduced correlator $\cT(U,V)$ which is our main focus in this paper.

The flat space limit formula \eqref{flat} can be applied to the conformal partial wave expansion in Mellin space
 \begin{align}
M(s,t)=\sum_{\ell=0}^{\infty}\int_{-i\infty}^{i\infty} d\nu b_\ell(\nu^2)M_{\nu,\ell}(s,t) \,,
\end{align}
where $M_{\nu,\ell}(s,t)$ are Mack polynomials and $b_\ell(\nu^2)$ has poles at the conformal dimensions of superconformal primaries with residues
\begin{align}
b_\ell(\nu^2)\approx \cC^2_{\tau,\ell} \frac{2^\ell K_{\Delta+4,\ell}}{\nu^2+(\Delta+4-2)^2}\,, 
\end{align}
where
\begin{align}
K_{\De,\ell}=\ & \frac{\Gamma(\Delta+\ell) \,\Gamma(\Delta-h+1) \, (\Delta-1)_\ell   }{ 
4^{\ell-1} 
\Gamma^4\!\left( \frac{\Delta +\ell}{2}\right)\Gamma^2\!\left( \frac{\Delta_i -\De+\ell}{2}\right)
\Gamma^2\!\left( \frac{\Delta_i +\De+\ell-d}{2}\right)
}\underset{\Delta\gg1}{\approx}\frac{2^{9+2\ell+2\Delta}}{\pi^3(\Delta)^{10+2\ell}}\sin^2\left(\frac{\pi\Delta}{2}\right).
\label{KDelta}
 \nonumber
\end{align}
 Following \cite{Costa:2012cb} the flat space limit of the conformal partial wave decomposition gives
 \beq
  \frac{f(s, t)}{s t u} = \sum\limits_{\ell=0}^\infty a_\ell(t) P_\ell(z)\,, \quad
  a_\ell(t) = \frac{L^6 c}{32} \left( \frac{L^2 t}{4} \right)^\ell b_\ell(-L^2 t)
  = \frac{\lambda^\frac{3}{2} (\a')^3 c}{32} \left( \frac{\sqrt{\lambda}\a' t}{4} \right)^\ell b_\ell(-\sqrt{\lambda}\a' t)\,,
 \eeq
 where $z=1+\frac{2s}{t}$ and $P_\ell(z)$ are partial waves for five-dimensional flat space ($C_\ell^{(\alpha)}(z)$ are Gegenbauer polynomials)
 \beq
 P_\ell(z) = 2^{-\ell} C_\ell^{(1)}(z)\,.
 \eeq
 Hence the OPE coefficients are given by
 \beq
\cC^2_{\tau,\ell} = - \frac{\sqrt{\lambda} \a'}{2^\ell K_{\De+4,\ell}} \Res\limits_{t=\frac{(\De+2)^2}{\sqrt{\lambda} \a'}} b_\ell\left(-\sqrt{\lambda}\a' t\right)\,.
 \label{ope_from_b}
 \eeq
We can now use the flat space result \eqref{flatspace_amplitude}
which has $t$-channel poles at $t=\frac{4\de}{\a'}$, $\de = 1,2,\ldots$ which are mapped to the residues in \eqref{ope_from_b} by the relation $\De = \lambda^\frac14 \sqrt{4 \de} - 2$. We can further use the orthogonality relation for Gegenbauer polynomials to compute the corresponding residues of the partial wave coefficients
\bea
\Res\limits_{t = \frac{4\de}{\a'}} a_\ell(t) &=
\int\limits_{-1}^1 dz
\frac{\sqrt{1-z^2}}{2^{-2\ell-1} \pi}
P_\ell(z) 
\Res\limits_{t = \frac{4\de}{\a'}}\frac{f(s, t)}{s t u} \\
&=
\int\limits_{-1}^1 dz
\frac{\sqrt{1-z^2}}{2^{-2\ell-1} \pi}
P_\ell(z) 
\frac{(\a')^2 (-1)^{\delta} \left(-\frac{1}{2} (1-z) \delta \right)_{\delta } \left(-\frac{1}{2} (1+z) \delta \right)_{\delta }}{4 \left(1-z^2\right) \delta^2 \Gamma (\delta+1)^2}\,.
\eea{a_from_flat_space}
Combining everything, the result for the OPE coefficients is
\beq
\cC^2_{\tau,\ell} = - \frac{ \pi^3 \tau^{10}}{c \lambda (\a')^2 2^{2\tau+3\ell+12} \sin^2(\frac{\pi \tau}{2})}
 \Res\limits_{t = \frac{4\de}{\a'}} a_\ell(t)\,.
\eeq
Using our definition \eqref{OPEStringy1}
\beq
\cC_{\tau,\ell}^2 = \frac{\pi^3 \tau^6}{c 2^{2\tau+2\ell + 12} (\ell+1)\sin^2(\frac{\pi \tau}{2})} f_0(\delta,\ell)\,,
\eeq
 we can also express the result as
\beq
f_0(\delta,\ell) = -  2^{4-\ell} \de^2 (\a')^{-2} (\ell+1) \Res\limits_{t = \frac{4\de}{\a'}} a_\ell(t)\,,
\eeq
and find formulas analytic in spin for the Regge trajectories labelled by $n = \delta-\frac{\ell}{2}-1 = 0, 1, 2, \ldots$
\bea
f_0(\tfrac{\ell}{2}+1,\ell) ={}& \frac{ (\ell+1) (\ell+2)^\ell}{2^{3 \ell} \Gamma \left(\frac{\ell}{2}+1\right)^2}\,,\\
f_0(\tfrac{\ell}{2}+2,\ell) ={}& \frac{ (\ell+1) (\ell+4)^{\ell+1} \left(\ell^2+11 \ell+12\right)}{3 \ 2^{3 \ell+6} \Gamma \left(\frac{\ell}{2}+2\right)^2}\,,\\
f_0(\tfrac{\ell}{2}+3,\ell) ={}& \frac{ (\ell+1) (\ell+6)^{\ell+1} \left(5 \ell^4+163 \ell^3+1870 \ell^2+7748 \ell+6360\right)}{45 \ 2^{3 \ell+12} \Gamma \left(\frac{\ell}{2}+2\right) \Gamma \left(\frac{\ell}{2}+3\right)}\,. 
\eea{f0}

 \section{Comparing to correlator of Lagrangians}
 \label{app:lagrangians}

In order to compare \eqref{f0} to \cite{Costa:2012cb,Goncalves:2014ffa} 
we need to relate the expansion coefficients $\cC^2_{\tau,\ell}$ in \eqref{Texp} to the OPE coefficients $\tl \cC^2_{\tau,\ell}$ in the four-point function where the external operators are the Lagrangian, which is a scalar operator of scaling dimension $\De=4$ which appears in the stress-tensor supermultiplet
 \begin{align}
&\left\langle \mathcal{L}(x_1) \mathcal{L}(x_2) \mathcal{L}(x_3) \mathcal{L}(x_4)  \right\rangle 
=\frac{G(U,V)}{x_{12}^8x_{34}^8} = \frac{1}{x_{12}^8x_{34}^8} \sum_{\tau,\ell}\tl \cC^2_{\tau,\ell}G_{\tau,\ell}(U,V)  \,.
\end{align}
The two correlators are related by supersymmetry and the relation is spelled out in \cite{Drummond:2006by,Goncalves:2014ffa}\footnote{In \cite{Goncalves:2014ffa} they use $F(U,V)=\frac{V}{U} \cT(U,V)$.} in terms of an eight-order differential operator in $U$ and $V$
\begin{align}
G(U,V) &= 2 \left( U^4 H\left(U,V\right) + H\left(1/U,V/U\right) + \frac{U^4}{V^4} H\left(U/V,1/V\right) \right)\,, \label{susy_relation}\\
H(U,V) &= \frac{1}{72} D^2 U^2 V^2 D^2 \frac{\cT\left(U,V\right)}{U^2}\,, \quad
D= U \partial_U^2 + V \partial_V^2 +(U+V-1)\partial_U \partial_V + 2 \partial_U+2\partial_V\,.\nonumber
\end{align}
We can use crossing symmetry \eqref{crossing_symmetry_T}
to obtain a differential operator $L$ directly relating the correlators of interest
\beq
G(U,V) = L \, \cT\left(U,V \right)\,.
\eeq
In order to study the effect of $L$ on \eqref{Texp}, we have to act on a generic conformal block and then expand the result in blocks. This can be made a lot easier by using a trick from \cite{Bissi:2019kkx}.
We assume $\frac{\tau}{2}-2,\ell \in \mathbb{N}$ and take a certain discontinuity of the block which turns it into an orthogonal polynomial
\beq
\underset{\zb\, \geq\, 1}{\Disc}\, \underset{z\, \geq\, 1}{\Disc}\, 
G_{4+2n,\ell}(z,\bar{z}) =
(2 \pi i)^2 r_{n+1} r_{n+\ell+2} \frac{ z \bar{z}}{z-\bar{z}}
P_{n,\ell}^{-}\left(\frac{2-z}{z}, \frac{2-\zb}{\zb} \right)\,, \quad r_n = \frac{\Gamma(2n)}{\Gamma(n)^2}\,,
\label{eq:DiscG_4d}
\eeq
where we defined $P_{n,\ell}^{-}$ as the antisymmetric two variable Legendre polynomial
\beq
P_{n,\ell}^{-} (x, \xb) = P_{n+\ell+1} ( x ) P_{n} ( \xb ) - P_{n} ( x ) P_{n+\ell+1} ( \xb )\,,
\qquad n, \ell = 0,1,2,\ldots \,,
\label{eq:P_def}
\eeq
which form a basis for antisymmetric two-variable polynomials. After acting with $L$, we use familiar relations between Legendre polynomials such as
\beq
x P_n(x)=\frac{n P_{n-1}(x)+(n+1) P_{n+1}(x)}{2 n+1}\,,
\eeq
to write everything as a sum of two-variable Legendre polynomials only, and read off the following relation, which we can check holds also for the full blocks for generic twist and spin
\begin{align}
\cT(U,V) ={}& U^{-2} G_{\tau+4,\ell} \Rightarrow\nonumber\\
G(U,V) ={}& \frac{1}{2^{13} 9} \Big( h_{0,0} G_{\tau ,\ell}
+h_{0,2}G_{\tau,\ell+2} +h_{0,4} G_{\tau ,\ell+4}+ h_{2,0} G_{\tau +2,\ell}+h_{2,2} G_{\tau +2,\ell+2}\nonumber\\
&+ h_{4,-2} G_{\tau +4,\ell-2}+h_{4,0} G_{\tau +4,\ell}+h_{4,2} G_{\tau+4,\ell+2}+h_{6,-2} G_{\tau +6,\ell-2} +h_{6,0} G_{\tau +6,\ell}\nonumber\\
&+h_{8,-4} G_{\tau +8,\ell-4}+h_{8,-4} G_{\tau +8,\ell-2}+h_{8,0} G_{\tau +8,\ell}   \Big)\,,\label{TtoG}
\end{align}
where
\begin{align}
h_{0,0}={}& 16 (\tau -6)^2 (\tau -4)^2 (2 \ell+\tau -4)^2 (2 \ell+\tau -2)^2\,,\nonumber\\
h_{0,2}={}& \frac{6 (\tau -6)^2 (\tau -4)^2 (2 \ell+\tau )^2 (2 \ell+\tau +2) (2 \ell+\tau +4) (2 \ell+\tau +6)^2 }{(2 \ell+\tau +1) (2 \ell+\tau +5)}\,,\nonumber\\
h_{0,4}={}& \frac{(\tau -6)^2 (\tau -4)^2 (2 \ell+\tau +4)^2 (2 \ell+\tau +6)^2 (2 \ell+\tau +8)^2 (2 \ell+\tau +10)^2 }{16 (2 \ell+\tau +3) (2 \ell+\tau +5)^2 (2 \ell+\tau +7)}\,,\nonumber\\
h_{2,0}={}& 16 (\tau -4)^2 (\tau -2) (\tau +2) (2 \ell+\tau ) (2 \ell+\tau +4) (2 \ell+\tau -2)^2\,, \nonumber\\
h_{2,2}={}& \frac{(\tau -4)^2 (\tau -2) (\tau +2) (2 \ell+\tau +2) (2 \ell+\tau +4)^2
   (2 \ell+\tau +6) (2 \ell+\tau +8)^2 }{(2 \ell+\tau +3) (2 \ell+\tau +5)}\,,\nonumber\\
h_{4,-2}={}& \frac{6 (\tau -2)^2 \tau  (\tau +2) (\tau +4)^2 (2 \ell+\tau -4)^2 (2 \ell+\tau   -2)^2 }{(\tau -1) (\tau +3)}\,, \nonumber\\
h_{4,0}={}& \frac{(\tau -2) \tau  (\tau +2) (\tau +4) (2 \ell+\tau ) (2
   \ell+\tau +2) (2 \ell+\tau +4) (2 \ell+\tau +6) }{4 (\tau -1) (\tau +3) (2 \ell+\tau +1) (2 \ell+\tau +5)} \Big(41 \tau ^4+328 \tau ^3\nonumber\\
   &+484 \tau ^2-688 \tau -480+4 \ell \left(41 \tau ^3+205 \tau ^2+78 \tau -504\right) + 4 \ell^2 \left(41 \tau ^2+82 \tau -168\right)\Big)\,, \nonumber
   \end{align}
   \begin{align}
h_{4,2}={}& \frac{3 (\tau -2)^2 \tau  (\tau +2)
   (\tau +4)^2 (2 \ell+\tau +4)^2 (2 \ell+\tau +6)^2 (2 \ell+\tau +8)^2 (2 \ell+\tau +10)^2 }{128 (\tau -1) (\tau +3) (2 \ell+\tau +3)
   (2 \ell+\tau +5)^2 (2 \ell+\tau +7)}\,, \nonumber\\
h_{6,-2}={}& \frac{\tau  (\tau +2)^2 (\tau +4) (\tau +6)^2 (2 \ell+\tau ) (2 \ell+\tau +4) (2 \ell+\tau-2)^2 }{\tau ^2+4 \tau +3}\,,\nonumber\\
h_{6,0}={}& \frac{\tau  (\tau +2)^2 (\tau +4) (\tau +6)^2 (2 \ell+\tau +2) (2 \ell+\tau +4)^2 (2 \ell+\tau +6) (2 \ell+\tau +8)^2 }{16 (\tau +1) (\tau +3) (2 \ell+\tau +3) (2 \ell+\tau +5)}\,,\nonumber\\
h_{8,-4}={}& \frac{(\tau +2)^2 (\tau +4)^2 (\tau +6)^2 (\tau +8)^2 (2 \ell+\tau -4)^2 (2 \ell+\tau -2)^2 }{16   (\tau +1) (\tau +3)^2 (\tau +5)}\,,\label{TtoG_coeffs}\\
h_{8,-2}={}& \frac{3 (\tau +2)^2 (\tau +4)^2 (\tau +6)^2 (\tau +8)^2 (2 \ell+\tau )^2 (2
   \ell+\tau +2) (2 \ell+\tau +4) (2 \ell+\tau +6)^2 }{128 (\tau +1) (\tau +3)^2 (\tau +5) (2 \ell+\tau +1) (2 \ell+\tau +5)}\,,\nonumber\\
h_{8,0}={}& \frac{\left((\tau +2) (\tau +4) (\tau +6) (\tau +8) (2 \ell+\tau +4) (2 \ell+\tau +6) (2 \ell+\tau +8) (2 \ell+\tau +10)\right)^2 }{4096 (\tau +1) (\tau +3)^2 (\tau +5) (2 \ell+\tau +3) (2 \ell+\tau +5)^2 (2 \ell+\tau +7)}\,.\nonumber
\end{align}
Note that whenever a spin on the right hand side of \eqref{TtoG} becomes negative (for $\ell=0$ or $\ell=2$) one has to replace
\beq
G_{\tau,\ell}(z,\zb) \to -G_{\tau+2\ell+2,-\ell-2}(z,\zb)\,,
\eeq
to obtain the correct expansion with non-negative spins.
At large $\tau$ \eqref{TtoG} becomes
\begin{align}
{}&\cT(U,V) = U^{-2} G_{\tau+4,\ell}(U,V) \Rightarrow\nonumber\\
{}&G(U,V)= \frac{\tau^8}{2^{13} 9} \bigg(
16 G_{\tau ,\ell}(z,\zb)+6 G_{\tau ,\ell+2}(z,\zb)+\frac{1}{16} G_{\tau ,\ell+4}(z,\zb)+16 G_{\tau +2,\ell}(z,\zb)\nonumber\\
&+G_{\tau
   +2,\ell+2}(z,\zb)+6 G_{\tau +4,\ell-2}(z,\zb)+\frac{41}{4} G_{\tau +4,\ell}(z,\zb)+\frac{3}{128} G_{\tau +4,\ell+2}(z,\zb)\label{rel_large_tau}\\
   &+G_{\tau
   +6,\ell-2}(z,\zb)+\frac{1}{16} G_{\tau +6,\ell}(z,\zb)+\frac{1}{16} G_{\tau +8,\ell-4}(z,\zb)+\frac{3}{128} G_{\tau
   +8,\ell-2}(z,\zb)+\frac{G_{\tau +8,\ell}(z,\zb)}{4096}\bigg)\,.\nonumber
\end{align}
This is the relation we will use to compare OPE coefficients of operators with large twist at strong coupling. It implies the following relation between OPE coefficients, where $n$ labels the Regge trajectories in the two correlators, starting from 0\footnote{Note that despite considering a large $\tau$ limit of the relation, we have to keep the finite shifts in $\tau$ in the arguments of OPE coefficients because they depend exponentially on $\tau$.}
\begin{align}
{}& \tl \cC_{n,J}^2 = \frac{\tau^8}{2^{13} 9} \bigg(
\frac{1}{16} \cC^2_{n,\tau ,J-4}
+\frac{3}{128} \cC^2_{n-1,\tau -4,J-2}+\cC^2_{n-1,\tau -2,J-2}+6 \cC^2_{n-1,\tau ,J-2}\nonumber\\
&-\frac{1}{16} \delta_{2,J}\cC^2_{n-1,\tau -2,0}
+\frac{\cC^2_{n-2,\tau -8,J}}{4096}+\frac{1}{16} \cC^2_{n-2,\tau
   -6,J}+\frac{41}{4} \cC^2_{n-2,\tau -4,J}+16 \cC^2_{n-2,\tau -2,J}\nonumber\\
   &+16 \cC^2_{n-2,\tau ,J}
   -\delta _{0,J} \left(\frac{3}{128} \cC^2_{n-2,\tau -6,0}+\cC^2_{n-2,\tau -4,0}+6 \cC^2_{n-2,\tau -2,0}\right)+\frac{3}{128} \cC^2_{n-3,\tau -8,J+2}\nonumber\\
&+\cC^2_{n-3,\tau -6,J+2}+6 \cC^2_{n-3,\tau -4,J+2}
   -\frac{1}{16} \delta _{0,J} \cC^2_{n-3,\tau -6,2}
   +\frac{1}{16} \cC^2_{n-4,\tau -8,J+4}
\bigg)\,. \label{ope_relation}
\end{align}
Here $J$ labels spins in the Lagrangian correlator and the corresponding twists for the OPE coefficients in both correlators are
\bea
\tl \cC_{n,J}^2 :&& \qquad \tau(J) &= \lambda^\frac{1}{4} 2 \sqrt{\frac{J}{2}+n-1} -J+2+ O(\lambda^{-\frac{1}{4}})\,,\\
\cC^2_{n,\tau,\ell} :&& \qquad \tau(\ell) &= \lambda^\frac{1}{4} 2 \sqrt{\frac{\ell}{2}+n+1} - \ell - 2 + O(\lambda^{-\frac{1}{4}})\,.
\eea{twists_C_lambda}
We are now ready to compare to \cite{Costa:2012cb,Goncalves:2014ffa} which computed the following OPE coefficients using the method reviewed in appendix \ref{app:f0}\footnote{To transcribe the formulas of \cite{Goncalves:2014ffa} to our conventions we had to multiply by $2^{-J}$ for the conformal block convention, $8$ for the normalisation of the correlator and $\pi^2$ (this factor does appear in \cite{Costa:2012cb}).
The expression for $\tl \cC_{1,J}^2$ in \cite{Goncalves:2014ffa} was wrong because they expanded the flat space amplitude in 10d partial waves instead of 5d partial waves.}
\bea
\tl \cC_{0,J}^2 &=\frac{\pi ^3  (J-2)^{J+5} \lambda ^{7/2}}{9 c \, 2^{2\tau+5 J+4} \Gamma \left(\frac{J}{2}\right)^2 \sin^2(\frac{\pi \tau}{2})}\,,\qquad  &&J = 4,6,8,\ldots\,, \\
\tl \cC_{1,J}^2 &=\frac{\pi ^3 J^{J+4} \left(J^2+87 J-16\right) \lambda ^{7/2} }{27 c \, 2^{2 \tau+5 J +8} \Gamma \left(\frac{J}{2}\right)^2 \sin^2(\frac{\pi \tau}{2})}\,, \quad &&J = 2,4,6,\ldots\,.
\eea{C_CRT}
Using \eqref{ope_relation} this precisely agrees with \eqref{f0}.

\bibliographystyle{JHEP}
\bibliography{paperbibliography}
\end{document}